\documentclass[10pt]{article}
\usepackage[english]{babel}
\usepackage[utf8x]{inputenc}
\usepackage[OT1]{fontenc}
\usepackage{amsfonts, amsmath, amsthm, amssymb, amsbsy}

\usepackage{subcaption}
\usepackage{multicol} 
\usepackage{stackrel}
\usepackage{graphicx}
\usepackage{listings}
\usepackage{csquotes}
\usepackage[hidelinks]{hyperref}
\usepackage[nottoc,notlot,notlof]{tocbibind}
\usepackage{cancel}
\usepackage[margin=0.9in]{geometry}
\usepackage{xcolor}
\usepackage{braket}

\title{An Exact Moment-Based Approach for Chemical Reaction-Diffusion Networks: From Mass Action to Hill Functions}

\author{
  \parbox{\linewidth}{ {\small
    Manuel Eduardo Hernández-García\textsuperscript{1,\dag}, Eduardo Moreno-Barbosa \textsuperscript{1, \S} and Jorge Velázquez-Castro\textsuperscript{1,\ddag}}\\
    {\footnotesize
    \textsuperscript{1}Facultad de Ciencias Físico Matemáticas, Benemérita Universidad Autónoma de Puebla, Heroica Puebla de Zaragoza 72570, México. \\
    \textsuperscript{\dag} \texttt{manuel.hernandezgarcia@viep.com.mx},  \textsuperscript{\S} \texttt{eduardo.morenoba@correo.buap.mx}, \textsuperscript{\ddag} \texttt{jorge.velazquezcastro@correo.buap.mx}.
  }}}

\date{\today}
\begin{document}
\maketitle 

\begin{abstract}
Biochemical systems are inherently stochastic, particularly those with small-molecule populations. The spatial distribution of molecules plays a critical role and requires the inclusion of spatial coordinates in their analysis. Stochastic models such as the chemical master equation are commonly used to study these systems. However, analytical solutions are limited to specific cases, and stochastic simulations require significant computational resources. To mitigate these challenges, approximation methods, such as the moment approach, reduce the system to a set of ordinary differential equations, thereby lowering the computational requirements. This study investigates the conditions under which the second-moment approach yields exact results during the dynamic evolution of chemical reaction-diffusion networks. The analysis encompasses second-order or higher-order reactions and Hill functions without relying on higher-order moment estimations or closure approximations. Starting with stationary states, we extended the analysis to a dynamic evolution. An enzymatic process and an antithetic feedback system were examined for purely reactive systems, demonstrating the approach's accuracy in capturing system behavior and quantifying errors. The study was further extended to genetic regulatory networks governed by Hill functions, including both purely reactive and reaction-diffusion systems, validating the method in spatially distributed contexts. This framework enables precise characterization of biochemical systems, avoiding information loss typically associated with approximations and allowing for stability analysis under fluctuations. These findings optimize computational strategies while providing insights into intracellular signaling and regulatory processes, paving the way for efficient and accurate stochastic modeling in biochemical systems.
\end{abstract}

\maketitle

\tableofcontents

\section{Introduction}
Biochemical reactions within cells are inherently subject to intrinsic fluctuations owing to the discrete nature of the system and the probabilistic occurrence of reactions \cite{Alon}. These fluctuations become particularly significant in systems with small-molecule populations, where randomness dominates dynamics \cite{Gillespie, Gar}. A stochastic description is more appropriate for such systems, offering deeper insights into the complex behavior of intracellular signaling and regulatory processes \cite{Ribeiro, VecchioM}. 
  
Moreover, molecules are also spatially distributed within cells, necessitating the inclusion of spatial coordinates in the analysis, which are also affected by fluctuations. One phenomenon emerging in reaction-diffusion systems is the formation of Turing patterns \cite{Turing}, a widely studied deterministic framework. However,  stochastic analysis of reaction-diffusion systems can reveal new regions where patterns emerge, providing insights that deterministic models may overlook \cite{Hori, Butler}. While deterministic models rely on systems of differential equations \cite{Turing, Manuel,  Shao, Scho}. Stochastic models, such as the Reaction-Diffusion Master Equation (RDME) \cite{Gar, Baya, Tha}, provide an exact probabilistic framework for describing system evolution. However, analytical solutions to the RDME are only tractable for simple systems, necessitating the use of numerical approaches, such as the Gillespie algorithm \cite{Gillespie}. Although these methods are also accurate, they are computationally expensive. Approximation techniques, including the Fokker-Planck equation \cite{Gar, Scott}, the Langevin equation \cite{VecchioM,  Hori, Langevin}, and the linear noise approximation \cite{Bianca}, are commonly used but remain computationally demanding and approximate the RDME. Moment-based approaches have been proposed as efficient alternatives, as shown in \cite{Manuel, Gomez}, which represent the system via a set of ordinary differential equations governing the central moments, allowing for the quantification of fluctuations throughout the dynamics of the system while significantly reducing the computational costs.
 
A specific application of the moment-based approach is the second-moment framework, where biochemical systems use the mean and second central moments (covariances) of concentrations and provide a simplified yet effective representation through ordinary differential equations \cite{ Manuel, Gomez}. However, it is well established that this approach is accurate only for zero- and first-order reactions \cite{Smad}.   A major challenge in moment-based methods, including second-moment framework, is the “closure problem,” which arises because the time evolution of moments up to order \(m\) depends on moments of order \(m+1\), particularly in the context of higher-order reactions \cite{Dowdy}.  

To overcome this problem, various closure schemes have been proposed. For example, the Zero-Information Closure Scheme \cite{Smad} estimates the  $m+1$-th moment from the $m$-th moment using principles from Shannon entropy. Similarly, the Kalman filter approach \cite{Ruess} estimates the  $m+1$-th moment from the $m$-th moment.  Additionally, approximations leveraging higher-order moments \cite{Ale} aim to improve accuracy. Another method, Multivariate Closure \cite{Lakatos}, assumes specific distributions for certain variables to derive higher-order moments and to close the system of equations. In addition, another method has been proposed in which effective parameters substitute high-order reactions; thus, the system is reduced to one with only zero- and first-order reactions \cite{Holimap}.  These methods provide viable solutions to the closure problem while balancing complexity and computational efficiency.

In this study, we investigated the closure problem in moment expansions for chemical reaction-diffusion networks that include spatial coordinates, extending previous works. Specifically, we focus on the conditions under which the second-moment framework can provide an exact description, even for second-order or higher-order reactions, and for Hill functions, which are frequently used in genetic regulatory networks \cite{Ribeiro, Shao, Schiavon, Kulasiri}, without requiring the estimation of higher-order moments or closure of the system of differential equations. This enables precise characterization of specific systems without the loss of information typically associated with approximations. Our analysis begins with stationary states, and is subsequently extended to examine the dynamic evolution of these systems.

Initially, we focused on systems that involve only chemical reactions. We began by analyzing an enzymatic process that reached a stationary state \cite{Grima} and then examined an antithetic feedback system that exhibited either a stationary point or a stable limit cycle depending on the parameter range \cite{Briat}. Additionally, we studied the dynamic evolution of these systems, quantifying the error throughout the process and demonstrating that the approach accurately captures their behavior. Subsequently, we extended our analysis to genetic regulatory networks governed by Hill functions, considering both purely reaction and reaction-diffusion systems. For these cases, we provided an exact description of their behavior and quantified the associated fluctuations, validating the approach in spatially distributed scenarios.\\

The remainder of this paper is organized as follows. In Section \ref{section2}, we derive the Reaction-Diffusion Master Equation. In Section \ref{section3}, we present a set of ordinary differential equations for the mean and second central moments, using the moment approach. In Section \ref{section4}, we identify the conditions under which this approach becomes exact. We also examined systems with higher-order reactions and provided equations for calculating the error at each time step for a specific set of differential equations describing the evolution of the chemical concentrations. In Section \ref{section5}, we analyze systems that do not include diffusion, focusing on enzymatic processes and antithetic integral feedback to assess the precision of the approximation for higher-order reactions. In addition, we analyzed a genetic regulatory network with negative feedback and described this system in detail. In Section \ref{section6}, we analyze a genetic regulatory network with negative feedback, similar to the case without diffusion, while incorporating diffusion. Finally, in Section \ref{section7}, we present our conclusions.

\section{Reaction-Diffusion Master Equation}  \label{section2}

In deterministic systems, the law of mass action provides a set of differential equations to describe concentration dynamics in chemical networks, incorporating diffusion terms for spatial effects. However, studying small systems with inherent fluctuations requires a stochastic approach, modeled using multivariable birth-death processes \cite{Gar}. In this context, the Reaction-Diffusion Master Equation plays a fundamental role, and its derivation begins with the introduction of essential definitions.   \\

\textbf{Definition 1 \cite{Anderson}.} A chemical reaction network is a triplet of non-empty, finite sets, usually denoted by $\mathcal{N}= \{\mathcal{S}, \mathcal{C}, \mathcal{R}\}$, where:
\begin{enumerate}     
\item A set of $N$  chemical species denoted by $\mathcal{S}= \{\mathcal{S}_1, \mathcal{S}_2, ..., \mathcal{S}_N\}$.      
\item A set of non-negative integer linear combinations of the species denoted by          

{\small
$\mathcal{C}=\{ \sum_{l=1}^{N} \alpha_{1l} \mathcal{S}_l, \sum_{l=1}^{N} \alpha_{2l} \mathcal{S}_l, ..., \sum_{l=1}^{N} \alpha_{Ml} \mathcal{S}_l, \sum_{l=1}^{N} \beta_{1l} \mathcal{S}_l, \sum_{l=1}^{N} \beta_{2l} \mathcal{S}_l, ..., \sum_{l=1}^{N} \beta_{Ml} \mathcal{S}_l \}$}, \\

the coefficients $\alpha_{il}$ and $\beta_{il}$ are non-negative integers, and they represent the stoichiometric coefficients.  
\item A set of $n$  chemical reactions denoted by $\mathcal{R}= \{\mathcal{R}_1, \mathcal{R}_2, ..., \mathcal{R}_n\}$. Through which these species are transformed, represented as ($i=(1,2,...,n)$)
\begin{align}      
\mathcal{R}_i : \sum_{l=1}^{N} \alpha_{il} \mathcal{S}_l \stackbin[]{k_i}{\rightarrow} \sum_{l=1}^{N} \beta_{il} \mathcal{S}_l. \label{1}
\end{align}
where $k_i$ is a parameter that denotes the rate of the reactions. The order of a reaction $\mathcal{R}_i$ is defined by $O(\mathcal{R}_i)=\sum_l \alpha_{il}$.
\end{enumerate}  

\begin{figure} [h!t]  
 \centering 
 \includegraphics[width=0.25\textwidth]{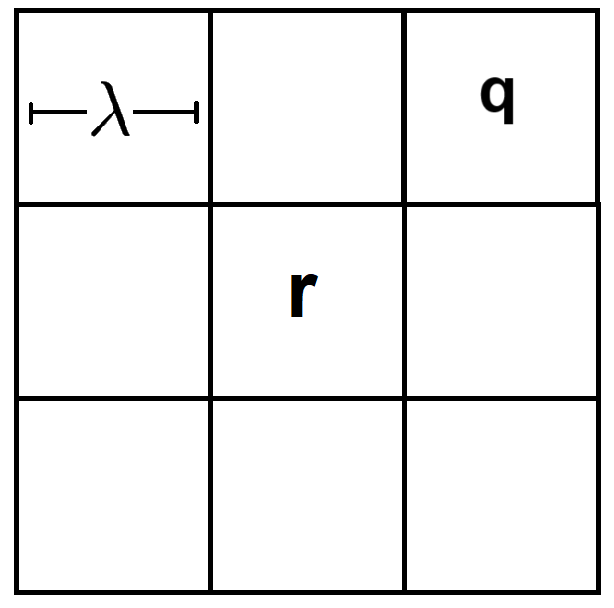}   \caption{ \textbf{Space domain partitioned into voxels.} The spatial domain is divided into voxels of uniform size, each with side length $\lambda$. Chemical species can diffuse from one voxel $r$ to another $q$, even if the voxels are not first neighborhoods.} \label{f.1} 
 \end{figure}

These definitions establish a framework for chemical reaction networks, focus on their reactions, and delineate the criteria that constitute a network of chemical reactions. However, the systems under consideration are spatially distributed. To model spatially distributed systems, we introduce a discretized domain \(\mathcal{J} \subset \mathbb{R}^3\), partitioned into finite regions termed voxels. A voxel represents the smallest spatial unit (e.g., a line segment, square, or cube) as illustrated in Figure \ref{f.1}. This discretization follows methodologies such as \cite{Hori, Lot}, with the alternative geometries discussed in \cite{Eng}. \\

 \textbf{Definition 2.} A chemical reaction-diffusion network is a non-empty set \(\mathcal{D}= \{ \mathcal{J}, \mathcal{N}^d \}\), where: 
 \begin{itemize}     
 \item \(\mathcal{J} \subset \mathbb{R}^3\) is a spatial domain discretized into \(J\) voxels of uniform characteristic size $\lambda$, adopting a geometry like the shown in Figure \ref{f.1}. The voxels do not overlap and cover the entire domain.
 
 \item Within each voxel \(r\) of \(\mathcal{J}\) (\(r \in \{1, 2, \dots, J\}\)), there exists a chemical reaction network, \(\mathcal{N}^r = \{\mathcal{S}^r, \mathcal{C}^r, \mathcal{R}^r\}\), where:    
 \begin{itemize}        
 \item \(\mathcal{S}^r\): the set of $N$ species in voxel \(r\), 
 \item \(\mathcal{C}^r\): the set of complexes in voxel \(r\),      
 \item \(\mathcal{R}^r\): the set of $n$ reactions in voxel \(r\).     
 \end{itemize} 
 From this, the set of chemical reaction networks over \(\mathcal{J}\) is denoted as \(\mathcal{N}^d = \{\mathcal{N}^1, \mathcal{N}^2, \dots, \mathcal{N}^J\}\).
 \item Let \(\mathcal{S}^r\) be a set of \(N\) chemical species in voxel \(r\),  which elements subsequently move to voxel \(q\). The diffusion process is expressed as follows,    
 \begin{align}         
 \mathcal{S}_{l}^r  \stackbin[]{d_{rq}^l}{\rightarrow} \mathcal{S}_{l}^q, \label{2}      
 \end{align} 
 where \(d_{rq}^l\) denotes the diffusion rate of the species \(\mathcal{S}_l^r\) ($\in \mathcal{S}^r$) from voxel \(r\) to voxel \(q\) (molecules per unit time) and and $d_{rr}^l =0$. 
 \end{itemize}

We assumed that all chemical species can be present in all voxels and that chemical reactions can occur in any voxel; therefore, the number of reactions and chemical species in each voxel is the same. With this framework, we now have a basis for modeling chemical reaction-diffusion networks. However, our focus is on stochastic systems or systems that incorporate intrinsic fluctuations in chemical species. Next, we constructed the Reaction-Diffusion Master Equation (RDME). 

 \subsection{Diffusion}
 
To construct the master equation associated with a diffusion process, we follow Definition 2, where there are $N$ different species in voxel $r$,  $S_l^r$ ($l \in \{1,2,..., N\}$),  and can move to voxel $q$, as illustrated in Figure \ref{f.1}. It is worth noting that the set of diffusion rates is directly related to the spatial discretization geometry so that the description can effectively capture all the movements of the chemical species between the cells. Thus, the scheme is independent of the chosen spatial partitioning \cite{Eng}.

Based on Definition 2, this process is described as follows: 
 \begin{align}   
 \mathcal{S}_{l}^r  \stackbin[]{d_{rq}^l}{\rightarrow} \mathcal{S}_{l}^q, \nonumber
 \end{align} 
the index $r$ indicates the initial region of the chemical species $l$, $q$ is the region to which it moves, and in general $d_{rq}^l \neq d_{qr}^l$ because the movement is not necessarily symmetric,  and $d_{rr}^l =0$. 

The diffusion process can be described in a similar way to a birth and death process but with propensity rates given as
 \begin{align}   
 \tau_{rq}^l=& d_{rq}^l \frac{S_{l}^r}{\Omega }, \label{3} 
 \end{align} 
where $S_l^r$ is the number of molecules of the chemical species $\mathcal{S}_l$ in voxel $r$,  $\Omega = N_{A}V$ ($V$ is the volume of each voxel \cite{Lot} ) is the size of the system and has units of $volume/mole$, and Avogadro's number $N_{A}$ is used to convert the number of the molecules to moles, and has units of $1/mol$ \cite{Decimal}. Using these propensity rates,  we obtained the master equation:

 {\footnotesize
 \begin{align}  
 \partial_{t} P(\mathbf{S}^r,t)= \Omega \sum_r \sum_l \sum_{q} & {d_{rq}^l} \left( \frac{S_{l}^r+1}{\Omega}  P(\mathbf{S}^r, S_l^r+1, S_{l}^q-1,t) - \frac{S_{l}^r}{\Omega}  P(\mathbf{S}^r,t)  \right). \label{4}
 \end{align} }
 We explicitly indicate which molecule moves from voxel $r$ to $q$, where $\mathbf{S}^r= (S_1^r,S_2^r,..., S_{N}^r)$ represents the state vector in the voxel $r$. The master equation describes the time evolution of the spatial distribution of molecules across the voxels, taking into account the inherent fluctuations. To simplify the representation, we express the master equation in terms of diffusion for each voxel $r$, $\mathrm{D}_r$. This allows the master equation to be compactly expressed as 
 \begin{align}      
 \partial_{t} P(\mathbf{S}^r,t)= {\Omega} \sum_r \mathrm{D}_r.  \label{5}
 \end{align}
A similar derivation can be found in \cite{Hori, Lot}, although with a different notation.

\subsection{Reactions}
Now, as the space is divided into voxels, we assume that each voxel is independent and that a chemical reaction network exists within each voxel. Following Definition 1, let be a voxel $r$, where there are $N$  chemical species $S_l^r$ ($l$ $\in$ \{$1,2,...,N$\}) and $n$ reactions $\mathcal{R}_i^r$ ($i$ $\in$ \{$1,2,...,n$\}), where the species is transformed as follows:   
 \begin{align}      
 \mathcal{R}_i^r : \sum_{l=1}^{N} \alpha_{il} S_l^r \stackbin[]{k_{i}^r}{\rightarrow} \sum_{l=1}^{N} \beta_{il} S_l^r. \nonumber 
 \end{align}  
 Coefficients $\alpha_{il} $ and $ \beta_{il}$ are non-negative integers. Note that we have considered that the parameters $k_{i}^r$ depend on the position, as in the following sections, we provide some examples of this case. From these expressions, we derived the stoichiometric matrix of the system, $ \Gamma_{il}= \beta_{li} -  \alpha_{li}$ (the indices are inverted to denote that it is the transpose). Through collisions (or interactions) between different elements, the system evolves according to the law of mass action and the propensity rates are given by  \cite{Gar}: 
 \begin{align}      
 a_i^r(\mathbf{S}^r)&= k_{i}^r \prod_{l} \frac{S_l^r !}{\Omega^{\alpha_{il}}(S_l^r- \alpha_{il} )!},   
 \end{align}   
 where index $i$ corresponds to the reactions $\mathcal{R}_i$ and $\mathbf{S}^r= (S_1^r, S_2^r, ..., S^r_{N})$. These propensities represent the transition probabilities between different system states per unit of time.  Given that reactions occur within each voxel, all voxels must be considered, yielding the following master equation: 

 {\small
 \begin{align}   
 \partial_{t} P(\mathbf{S}^r,t)= \Omega \sum_r \sum_{i=1}^{n} \left( a_{i}^r(\mathbf{S}^r-\Gamma_{i}) P(\mathbf{S}^r-\Gamma_{i}^{r},t) - a_{i}^r(\mathbf{S}^r) P(\mathbf{S}^r,t) \right), \label{7}  
 \end{align}}   
 where $\Gamma_{i}$ represents the $i$-th column of matrix $\Gamma$. We can rewrite the above master equation in a more compact form by representing it as a sum of the reactions $\mathrm{R}_r$ in each voxel $r$, leading to 
 \begin{align}       
 \partial_{t} P(\mathbf{S}^r,t)= {\Omega} \sum_r \mathrm{R}_r. \label{8} 
 \end{align}  
Assuming a system in which fluctuations occur in both reactions and diffusion, we combine the respective master equations: Equation (\ref{5}) for reactions and Equation (\ref{8}) for diffusion. This yields a unified master equation that accounts for both effects:
 \begin{align}    
 \partial_{t} P(\mathbf{S}^r,t) = {\Omega} \sum_r \left( \mathrm{R}_r + \mathrm{D}_r \right). \label{9} 
 \end{align} 
 The master equation describes the temporal evolution of the probability distribution of molecules subjected to the reaction and diffusion phenomena. It is important to note that, at this stage, no approximations regarding the spatial distribution of the molecules have been made; this aspect will be addressed later.  However, directly solving  Equation (\ref{9}) requires significant computational power; thus, approximations might be employed. However, obtaining exact results for summary metrics, such as the central moments, is desirable, as this reduces the computational cost of the description without sacrificing accuracy.

\section{Moment Approach }\label{section3} 
Solving the master equation (\ref{9}) directly is computationally expensive even when standard numerical methods are employed \cite{Gillespie}. Therefore, approximation methods are required to describe these dynamics. In this study, we use moment expansion  \cite{Lakatos}, which transforms the master equation into a set of ordinary differential equations for the central moments. Specifically, we consider up to the second central moment that allows us to quantify the fluctuations of the system \cite{Manuel, Gomez}. It is important to note that higher-order central moments can also be used \cite{Lakatos}, although the second central moment is typically sufficient to capture the essential features of fluctuations in many cases.\\

First, we define the following quantities:
\begin{itemize}
    \item The mean concentration of chemical species in the voxel $r$: 
     \begin{align}
         s_l^r = \frac{\langle S_l^r \rangle}{\Omega}.
     \end{align}
    \item The $m$-th central moments of the chemical species between different voxels are:
    
    {\small
\begin{align}
    M^m_{l_1^{r_1},l_2^{r_2}, ... , l_m^{r_m}}= \frac{\braket{(S_{l_1}^{r_1}- \braket{S_{l_1}^{r_1}})(S_{l_2}^{r_2}- \braket{S_{l_2}^{r_2}})... (S_{l_m}^{r_m}- \braket{S_{l_m}^{r_m}})}}{\Omega^m} .
\end{align}}
\end{itemize}

As previously mentioned, $\Omega$ helps convert molecules to concentrations.

In this work, we are going to consider analytic functions  $f(\mathbf{S}^{r_1}, \mathbf{S}^{r_2},..., \mathbf{S}^{r_R})$ (where $\mathbf{S}^r=(S_1^r, S_2^r,..., S_{N}^r)$) of the system variables that can be expanded using a Taylor expansion around the mean, as shown below:

{\small
\begin{align}
   \braket{f(\mathbf{S}^{r_1}, \mathbf{S}^{r_2},..., \mathbf{S}^{r_R})} =& \left \langle f( \braket{\mathbf{S}^{r_1}}, \braket{\mathbf{S}^{r_2}},..., \braket{\mathbf{S}^{r_R}}) + \sum_{a_1=1}^R \sum_{j_1} (S_{j_1}^{r_{a_1}}- \braket{S_{j_1}^{r_{a_1}}})\frac{\partial f( \braket{\mathbf{S}^{r_1}}, \braket{\mathbf{S}^{r_2}},..., \braket{\mathbf{S}^{r_R}})}{\partial S_{j_1}^{r_{a_1}}} \right. \nonumber \\
   +& \left. \sum_{m=2}^{\infty} \sum_{a_1,a_2, ... , a_n} \sum_{j_1,j_2, ... , j_m} \frac{(S_{j_1}^{r_{a_1}}- \braket{S_{j_1}^{r_{a_1}}})(S_{j_2}^{r_{a_2}}- \braket{S_{j_2}^{r_{a_2}}})... (S_{j_m}^{r_{a_m}}- \braket{S_{j_m}^{r_{a_m}}})}{m!} \frac{\partial^m f( \braket{\mathbf{S}^{r_1}}, \braket{\mathbf{S}^{r_2}},..., \braket{\mathbf{S}^{r_R}})}{ \partial S_{j_1}^{r_{a_1}} \partial S_{j_2}^{r_{a_2}}... \partial S_{j_m}^{r_{a_m}}} \right \rangle \nonumber \\ 
   =&f( \braket{\mathbf{S}^{r_1}_1}, \braket{\mathbf{S}^{r_2}_2},..., \braket{\mathbf{S}^{r_n}_n}) +  \sum_{m=2}^{\infty} \sum_{a_1,a_2, ... , a_m} \sum_{j_1,j_2, ... , j_m} \frac{ C^m_{j_1^{r_{a_1}},j_2^{r_{a_2}}, ... , j_m^{r_{a_m}}}}{m!} \frac{\partial^m f( \braket{\mathbf{S}^{r_1}}, \braket{\mathbf{S}^{r_2}},..., \braket{\mathbf{S}^{r_R}})}{ \partial S_{j_1}^{r_{a_1}} \partial S_{j_2}^{r_{a_2}}... \partial S_{j_m}^{r_{a_m}}}, \label{12} 
\end{align}}

where $\braket{\mathbf{S}^r}= (\braket{S_1^r},\braket{S_2^r},...,\braket{ S_{N}^r})$  is the mean state vector on voxel $r$, and $C^m_{j_1^{r_1},j_2^{r_2}, ... , j_m^{r_m}} = \Omega^m M^m_{j_1^{r_1},j_2^{r_2}, ... , j_m^{r_m}}$.  If function $f$ is a polynomial of order $a$ ($a \in \mathbb{N}$), then the expansion in Equation (\ref{12}) terminates at the $a$-th order central moment expansion.

By multiplying $S_{l}^{r}$ by (\ref{12}) and integrating, we obtain the system of ordinary differential equations (\ref{13}) for the dynamical evolution of the mean concentrations. Similarly, an equation describing the evolution of the second moments (\ref{14}) can be obtained. In equations (\ref{13}) and (\ref{14}), the expansion (\ref{12}) was used.

{\small
\begin{align}     
\frac{\partial s_l^r }{\partial{t}} & =  \sum_{i} \Gamma_{li} k_i^r \left( R_i(\mathbf{s}^r) + \sum_{m=2}^{\infty} \sum_{j_1,j_2, ... , j_m} \frac{M^m_{j_1^r,j_2^r, ... , j_m^r}}{m!} \frac{\partial^m R_i(\mathbf{s}^r)}{ \partial s_{l_1}^r \partial s_{l_2}^r... \partial s_{l_m}^r} \right) +  \sum_q  \left(  d_{qr}^l {s_l^q} - d_{rq}^l {s_l^r} \right), \label{13} \\
\frac{\partial M^2_{{l_1}^{r_1}, {l_2}^{r_2}}}{\partial{t}}  & = \sum_{i} \left( \delta_{r_1,r_2} \frac{\Gamma_{l_1 i} \Gamma_{l_2 i}}{\Omega} k_i^{r_1} \left( R_i(\mathbf{s}^{r_1}) + \sum_{m=2}^{\infty} \sum_{j_1,j_2, ... , j_m} M^m_{j_1^{r_1},j_2^{r_1}, ... , j_m^{r_1}} \frac{\partial^m R_i(\mathbf{s}^{r_1})}{ \partial s_{j_1}^{r_1} \partial s_{j_2}^{r_1}... \partial s_{j_m}^{r_1}}  \right) \right. \nonumber \\     
& +  \sum_{j_1=1}^{N} \left( M^2_{{l_1}^{r_1}, {j_1}^{r_2}} \Gamma_{l_2 i} k_i^{r_2} \frac{\partial R_i(\mathbf{s}^{r_2})}{\partial {s_{j_1}^{r_2}}}  + M^2_{{j_1}^{r_1}, {l_2}^{r_2}} \Gamma_{l_1 i} k_i^{r_1} \frac{\partial R_i(\mathbf{s}^{r_1})}{\partial {s_{j_1}^{r_1}}} \right) \nonumber \\ 
&+ \left. \sum_{m=2}^{\infty} \sum_{j_1,j_2, ... , j_m} \left(M^{m+1}_{j_1^{r_2},j_2^{r_2}, ... , j_{m}^{r_2}, l_{1}^{r_1}} \Gamma_{l_2 i} k_i^{r_2} \frac{\partial^m R_i^{r_2}(\mathbf{s}^{r_2})}{ \partial s_{j_1}^{r_2} \partial s_{j_2}^{r_2}... \partial s_{j_m}^{r_2}} + M^{m+1}_{j_1^{r_1},j_2^{r_1}, ... , j_{m}^{r_1}, l_2^{r_2}} \Gamma_{l_1 i} k_i^{r_1} \frac{\partial^m R_i^{r_1}(\mathbf{s}^{r_1})}{ \partial s_{j_1}^{r_1} \partial s_{j_2}^{r_1}... \partial s_{j_m}^{r_1}} \right) \right) \nonumber \\
&+  \frac{\delta_{l_1, l_2}}{\Omega} \left( \delta_{r_1,r_2} \sum_q \left( d_{s r_1}^{l_1} s_{l_1}^{q} + d_{r_1 q}^{l_1} s_{l_1}^{r_1} \right) -(d_{r_1 r_2}^{l_1} s_{l_1}^{r_1} + d_{r_2 r_1}^{l_1} s_{l_1}^{r_2}) \right) \nonumber \\
&+  \sum_q \left( d_{qr_1}^{l_1} M^2_{l_1^{q},l_2^{r_2}}  - d_{r_1q}^{l_1} M^2_{l_1^{r_1},l_2^{r_2}} \right) +  \sum_q \left( d_{qr_2}^{l_2} M^2_{l_1^{r_1},l_2^{q}}  -d_{r_2q}^{l_2} M^2_{l_1^{r_1},l_2^{r_2}}  \right),  \label{14}
\end{align}}

\noindent here $\mathbf{s}^r=(s_1^r, s_2^r,..., s_{N}^r)$), $\delta_{r_1,r_2}$ is the Kronecker delta, and $R_i(\mathbf{s}) =  \prod_{j=1}^{N} \prod_{z=1}^{\alpha_{ij}} (s_j^r - \frac{\alpha_{ij}-z}{\Omega})$ are the reaction rates.  

It is worth noting that equations (\ref{13}) and (\ref{14}) are still exact since no approximation has been made in the spatial variables; refer to \cite{Lot} for a continuous approximation of spatial coordinates.

This approach facilitates the development of computational programs and ensures that the equations are generalizable to other types of models like epidemics on metapopulation network \cite{GIMENEZ} or cellular automata \cite{Ila}. Using this system of ordinary differential equations, it is possible to simultaneously determine the dynamics of concentrations and their spatial distributions. Additionally, stochastic corrections due to reactions and diffusion across regions were incorporated, enabling the quantification of fluctuations throughout the time evolution of the system. Comparable approximated expressions were reported in \cite{Lot}, where the derivation was based on linear noise approximation.

In general, equations (\ref{13}) and (\ref{14}) are coupled with higher order moments other than the second central moment, leading to what is known as the closure problem. To address the closure problem, we analyze the conditions under which the differential equations in (\ref{13}) and (\ref{14}) remain exact without requiring third or higher central moments, thereby simplifying the system while retaining its accuracy.

\section{Exact Approach for Chemical Reaction-Diffusion Network} \label{section4}

In this section, we identify the conditions under which an approach up to the second moment becomes exact, even in systems involving second- or higher-order reactions and Hill-type functions, while accounting for diffusion. This is particularly significant because it enables precise analysis of a broader class of systems throughout their dynamic evolution. Additionally, this approach is particularly valuable when studying stationary states because it facilitates stability analysis in the presence of fluctuations. Here, we define the specific criteria under which a stochastic system can be accurately described using only the mean and second central moments as captured in Equations (\ref{13}) and (\ref{14}), respectively. These conditions ensure that the system dynamics and fluctuations are faithfully represented, thereby eliminating the need to calculate higher-order moments in these scenarios. 
 
First, we provide definitions for classifying the types of chemical reaction-diffusion networks.  \\
 
 \textbf{Definition 3.} A chemical reaction-diffusion network \(\mathcal{D}\) is classified as follows:   
 \begin{itemize}     
 \item It is \textbf{mass-action kinetic with diffusion} if all parameters \(k_i^r\) and \(d_{rs}^l\) do not exhibit fluctuations.  
 \item It is \textbf{non-mass-action kinetic with diffusion} if the  parameters \(k_i^r\) have a functional form that depends on the mean concentrations and central moments       \[     k_i^r = k_i^{*r} f_i(\mathbf{s}^r, \mathbf{M}^2_{r,r}, \ldots, \mathbf{M}_{r,r,...,r}^m),  \] 
 and if all parameters \(k_i^{*r}\) and \(d_{rs}^l\) do not exhibit fluctuations. \(k_i^{*r}\) is a constant parameter. 
 \end{itemize}

The idea of a functional parameter was derived from the work developed in \cite{Manuel}, in which the derivation of the Hill function from a stochastic framework makes it equivalent to a functional parameter that depends on the mean concentration and second central moment. However, this idea has also been explored in a previous study \cite{Holimap} in which they used effective parameters and the closure problem does not appear.  When parameters fluctuate, as discussed in a previous work \cite{Extri}, such systems are referred to as having extrinsic fluctuations. However, in this study, we focus exclusively on systems in which the parameters remain constant and do not exhibit fluctuations. Consequently, this study quantifies the effects of intrinsic fluctuations and determines the specific conditions under which the second-moment approach is exact, even for non-mass-action propensity functions, such as Michaelis-Menten or Hill functions, that explicitly depend on moments, as demonstrated in \cite{Manuel}, where the Hill equation depends on mean concentration and central moments as it captures fast reactions.  The Hill functions are frequently used \cite{Schnoerr}, particularly in Genetic Regulatory Networks \cite{Ribeiro, Shao, Schiavon, Kulasiri}. For a detailed explanation of this dependence and its derivation, please refer to Appendix \ref{C}. This approach highlights the versatility of the moment-based approach, allowing it to accommodate a broad spectrum of reaction dynamics, extending beyond the constraints of mass-action kinetics.  \\

\textbf{Proposition 1. \label{P.1}}  
Let \(\mathcal{D}\) be a chemical reaction-diffusion network, where:
\begin{itemize}
    \item There are only zero- and first-order reactions in each voxel \(r\).
    \item The parameters have a functional form that depends on the mean concentrations $\mathbf{s}^r$ and $m$-th central moments \(\mathbf{M}_{r,r,...,r}^m\), such that \[     k_i^r = k_i^{*r} f_i(\mathbf{s}^r, \mathbf{M}^2_{r,r}, \ldots, \mathbf{M}_{r,r,...,r}^m).   \] 
\end{itemize}
Then, the system can be described exactly using the time differential equations up to the \(m\)-th central moment. \\

In Appendix \ref{A}, we present the equations for the dynamics and demonstrate the proposition.  Although our primary objective is to determine the conditions under which a stochastic system can be described exactly up to the second central moment, we present the previous proposition that extends the framework to systems in which the effective parameters depend on higher-order moments. This generalization allows us to model more complex systems with intricate dependencies. In the following sections, we provide an example to illustrate this scenario in detail. It is essential to emphasize that these equations remain exact at any time $t$, including in the stationary state.   

Based on this proposition, we derive a corollary for systems where the effective parameters depend only on the second central moment. This corollary simplifies the analysis of such systems while maintaining their accuracy, making it a valuable tool for studying a wide range of stochastic dynamics.  \\

\textbf{ Corollary 1. \label{C.1}} 
Let \(\mathcal{D}\) be a chemical reaction-diffusion network, where:
\begin{itemize}
    \item There are only zero- and first-order reactions in each voxel \(r\).
    \item The parameters have a functional form that depends on the mean concentrations $\mathbf{s}^r$ and second central moments \(\mathbf{M}_{r,r}^2\), such that \[     k_i^r = k_i^{*r} f_i(\mathbf{s}^r, \mathbf{M}^2_{r,r}).   \] 
\end{itemize}
The system can then be described using the following differential equations:

\begin{align}     
\frac{\partial s_l^r }{\partial{t}} & =  \sum_{i} \Gamma_{li} k_i^r R_i(\mathbf{s}^r)   +  \sum_q  \left(  d_{qr}^{l} {s_l^q} - d_{rq}^{l} {s_l^r} \right), \nonumber \\
\frac{\partial M^2_{{l_1}^{r_1}, {l_2}^{r_2}}}{\partial{t}}  & = \sum_{i} \left( \delta_{r_1,r_2} \frac{\Gamma_{l_1 i} \Gamma_{l_2 i}}{\Omega} \left(k_i^{r_1} R_i(\mathbf{s}^{r_1})   \right) \right. \nonumber \\
&+ \left. \sum_{j_1=1}^{N} \left( M^2_{{l_1}^{r_1}, {j_1}^{r_2}} \Gamma_{l_2 i} k_i^{r_2} \frac{\partial R_i(\mathbf{s}^{r_2})}{\partial {s_{j_1}^{r_2}}}  + M^2_{{j_1}^{r_1}, {l_2}^{r_2}} \Gamma_{l_1 i} k_i^{r_1} \frac{\partial R_i(\mathbf{s}^{r_1})}{\partial {s_{j_1}^{r_1}}} \right)   \right) \nonumber \\
&+  \frac{\delta_{l_1, l_2}}{\Omega} \left( \delta_{r_1,r_2} \sum_q \left( d_{q r_1}^{l_1} s_{l_1}^{q} + d_{r_1 q}^{l_1} s_{l_1}^{r_1} \right) -(d_{r_1 r_2}^{l_1} s_{l_1}^{r_1} + d_{r_2 r_1}^{l_1} s_{l_1}^{r_2}) \right) \nonumber \\
&+  \sum_q \left( d_{qr_1}^{l_1} M^2_{l_1^{q},l_2^{r_2}}  - d_{r_1q}^{l_1} M^2_{l_1^{r_1},l_2^{r_2}} \right) +  \sum_q \left( d_{qr_2}^{l_2} M^2_{l_1^{r_1},l_2^{q}}  -d_{r_2q}^{l_2} M^2_{l_1^{r_1},l_2^{r_2}}  \right),  \label{15}
\end{align}

where $R_i(\mathbf{s}^{r})= \prod_j (s_j^r)^{\alpha_{ij}}$. \\

A heuristic way to show this affirmation is to observe that in Equations (\ref{13}) and (\ref{14}), the derivatives of the reaction rates beyond the second order are zero because the reaction rates are, at most, linear functions. As a result, these equations remain exact at any time $t$, including the stationary state. When all effective parameters do not depend on moments. The equations presented in Equation (\ref{15}) coincide with the linear noise approximation \cite{Vanka, Lot} when the parameters follow mass-action kinetics; however, in Proposition 1, we generalize the conditions in which these are exact. 
 
In the previous proposition and its corollary, we outlined the conditions under which an exact approach can be achieved for zero- and first-order reactions. However, it is equally critical to extend this analysis to include higher-order reactions because such reactions frequently occur in various biochemical systems \cite{Grima}. Before delving into this extension, we must establish the following propositions to formalize the conditions and framework for achieving an exact approach in systems involving higher-order reactions. \\

\textbf{Proposition 2.} Let $\mathcal{D}$ be a chemical reaction-diffusion network like in Definition 2. The set of chemical species in each voxel $r$ is denoted as $\mathcal{S}^r$. If the rank of the stoichiometric matrix $\Gamma$ in voxel $r$ equals the number of chemical species in that voxel, $N$, that is, \(\text{Rank}(\Gamma) = N\), then there are no conserved quantities by stoichiometric reactions. \\  

Note that we assume a chemical reaction-diffusion network in Proposition 2, as presented in Definition 2; therefore, the number of different chemical species is the same in each voxel. Another important observation is that, because there are no conserved quantities in the dynamics, the chemical species are not dependent on conservative quantities \cite{Klamt}.  \\

\textbf{Proposition 3.} Let \(\mathcal{D}\) be a chemical reaction-diffusion network like on Definition 2, and let \(\mathcal{S}^{r}\) be the set of chemical species in voxel \(r\). If in the stationary state, the chemical species are uncorrelated, and the central moments satisfy the following conditions: 
\begin{align}     
M^{m}_{j_1^{r},j_2^{r}, ... , j_m^{r} }=   
\begin{cases}         
\neq 0 & \text{if } m \geq 2 \text{ and } j_1 = j_2= ...=j_m \\ 
0 & \text{other cases}      
\end{cases} . 
\end{align} 
Then the set of chemical species in voxel \(r\) is independent. \\   

This proposition is inspired by the work of Grima \cite{Grima}, where at least one chemical species needs to be uncorrelated with the others. \\

\textbf{Proposition 4.} Let \(\mathcal{D}\) be a chemical reaction-diffusion network like on Definition 2. For two arbitrary sets of independent chemical species, \(\mathcal{S}_1^{r_1}\) and \(\mathcal{S}_2^{r_2}\), in voxels \(r_1\) and \(r_2\), respectively. If the set of variables \(\mathcal{S}_{1,2} = \{ \mathcal{S}_1^{r_1}, \mathcal{S}_2^{r_2}\}\) is also independent, then the chemical species are uncorrelated in the stationary state and the central moments satisfy the following conditions:
{\small
\begin{align}     
M^{m+1}_{j_1^{r_1},j_2^{r_1}, ... , j_m^{r_1},l_2^{r_2} }=      
\begin{cases}         
\neq 0 & \text{if } m \geq 1 \text{ and } j_1 = j_2= ...=j_m=l_2 \\        
0 & \text{other cases}      
\end{cases} . \label{16}
\end{align}}  

The last three propositions establish the conditions under which the species in the system are independent and uncorrelated, a key ingredient that will be used in the next proposition. \\

\textbf{Proposition 5. \label{P.5}}  
Let \(\mathcal{D}\) be a chemical reaction-diffusion network, where in each voxel \(r\):
\begin{itemize}
    \item The stoichiometric coefficients $\alpha_{ij}$ satisfy: \(\alpha_{ij} \leq 1\),
    \item The set of chemical species \(\mathcal{S}^r\) is independent.
    \item The parameters have a functional form that depends on the mean concentrations $\mathbf{s}^r$ and second central moments \(\mathbf{M}_{r,r}^2\), such that \[     k_i^r = k_i^{*r} f_i(\mathbf{s}^r, \mathbf{M}^2_{r,r}).   \] 
\end{itemize}
Additionally, for any two distinct voxels, \(r_1\) and \(r_2\), each satisfies the above conditions. Let \(\mathcal{S}_1^{r_1}\) and \(\mathcal{S}_2^{r_2}\) denote two sets of independent chemical species for each voxel. If the combined set \(\mathcal{S}_{1,2} = \{ \mathcal{S}_1^{r_1}, \mathcal{S}_2^{r_2}\}\) is also independent, then at a steady state, the system can be described exactly using only the following equations:

\begin{align}     
\frac{\partial s_l^r }{\partial{t}}= 0 & =  \sum_{i} \Gamma_{li}  k_i^r R_i(\mathbf{s}^r)  +  \sum_q  \left(  d_{qr}^{l} {s_l^q} - d_{rq}^{l} {s_l^r} \right) ,\nonumber \\
\frac{\partial M^2_{{l}^{r_1}, {l}^{r_2}}}{\partial{t}}= 0  & = \sum_{i} \left( \delta_{r_1,r_2} \frac{\Gamma_{l i} \Gamma_{l i}}{\Omega} k_i^{r_1} R_i(\mathbf{s}^{r_1})  \right. \nonumber \\ 
& +  \left.  \left( M^2_{{l}^{r_1}, {l}^{r_2}} \Gamma_{l i}  k_i^{r_2} \frac{\partial R_i(\mathbf{s}^{r_2})}   {\partial {s_{l}^{r_2}}}  + M^2_{{l}^{r_1}, {l}^{r_2}} \Gamma_{l i}k_i^{r_1} \frac{\partial R_i(\mathbf{s}^{r_1})}{\partial {s_{l}^{r_1}}} \right)  \right) \nonumber \\
&+  \frac{1}{\Omega} \left( \delta_{r_1,r_2} \sum_q \left( d_{q r_1}^{l} s_{l}^{q} + d_{r_1 q}^{l} s_{l}^{r_1} \right) -(d_{r_1 r_2}^{l} s_{l}^{r_1} + d_{r_2 r_1}^{l} s_{l}^{r_2}) \right) \nonumber \\
&+  \sum_q \left( d_{q{r_1}}^{l} M^2_{l^{q},l^{r_2}}  - d_{{r_1}q}^{l} M^2_{l^{r_1},l^{r_2}} \right) +  \sum_q \left( d_{sr_2}^{l} M^2_{l^{r_1},l^{q}}  -d_{r_2s}^{l} M^2_{l^{r_1},l^{r_2}}  \right),  \label{17}
\end{align}

where $R_i^{r}(\mathbf{s})= \prod_j (s_j^r)^{\alpha_{ij}}$ and the other second central moments are zero. \\ 

The proof of this proposition is provided in Appendix \ref{B}. If dependent chemical species are present, the system can be reduced to an independent set of variables, and the dependent variables can be substituted into the reaction rates. If the degree of each variable involved in these reaction rates remains of order one, we can still obtain an exact description analogous to Equation (\ref{17}).   An important observation of this proposition is that, to obtain an accurate description, the chemical species must appear at most once in the reactions of the system. Another important component is that they must be independent, which leads to them being uncorrelated, as in \cite{Grima}.  The significance of this proposition lies in its potential to facilitate the development of stability-analysis methods. 

Motivated by Proposition 5, we investigate the accuracy of the second-moment approach for similar systems but consider the entire dynamical evolution. To achieve this, we modeled the system using Equation (\ref{15}) only until the second central moment. In the subsequent analysis, we employed various magnitudes of $\Omega$ to compare the results obtained from the stochastic simulations using the Gillespie Algorithm \cite{Gillespie} and the values of the mean concentration (second central moment) over 10,000 simulations ($s_{j, SS}(t)$) with those derived using (\ref{15}) ($s_{j, Ap}(t)$). To accomplish this, we will calculate the error at each time point as following
\begin{align}
    E_j(t)= \frac{1}{t} \int_0^t (s_{j,SS}(t')- s_{j,Ap}(t'))^2 dt'.  \label{20}
\end{align}

The results we obtained were for reaction-diffusion systems; however, if diffusion is removed, these results are still valid for chemical reaction networks. Next, we analyze some particular systems for applying the formalism developed in this section. Focus on the stationary and dynamic evolution of systems.  

\section{Chemical Reaction Network} \label{section5}
In the first part, we analyzed three different systems to illustrate the application of the propositions and corollaries presented in the previous section, where we took the diffusion parameters equal to zero. This analysis provides insights into the behavior of reaction-only systems and how the second-moment approach can be used to accurately describe fluctuations within these systems.  

These characteristics make moment-based methods particularly valuable for analyzing and modeling the dynamics of these systems. Now we give conditions under the second-moment approach to provide an exact description for mass action kinetic with diffusion, this type of system appears in a variety of contexts, such as enzyme kinetics \cite{Lower, Cooney}, or protein interactions \cite{Durham}, where reactions occur according to mass action laws, and diffusion governs the movement of molecules within cells or tissues.

\subsection{Enzyme Processes} 

\begin{figure} [h!t]
\centering
\includegraphics[width=.3\textwidth]{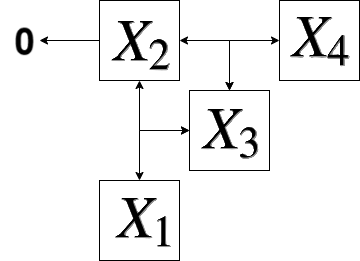}\hfill
\caption{\textbf{Enzymatic Processes.} In this figure, there is a representation of the enzymatic processes, in which molecule $X_1$ binds $X_2$ in a reversible process to form complex $X_3$, where this complex then produces product $X_4$ and releases $X_2$ in a reversible process.}
\label{fig.0}
\end{figure}

The first system that we analyzed was an enzymatic process \cite{Grima}. As shown in Figure \ref{fig.0}, enzymes play a crucial role in biological functions by acting as catalysts to accelerate chemical reactions in living organisms \cite{Ainsworth}. They are involved in various processes including replication, transcription, protein synthesis, metabolism, and signaling \cite{Proud}. 

The system is described by the following reactions,
\begin{table}[h!]
    \centering
    \begin{tabular}{|c|c|}
        $\emptyset  \stackbin{k_1}{\longrightarrow} X_2$ & $X_2  \stackbin{k_2}{\longrightarrow} \emptyset$ \\
        $X_1 + X_2  \stackbin{k_3}{\longrightarrow} X_3 $ & $ X_3  \stackbin{k_4}{\longrightarrow} X_1 + X_2 $ \\
        $X_3  \stackbin{k_5}{\longrightarrow} X_2 + X_4$ & $ X_2 + X_4 \stackbin{k_6}{\longrightarrow} X_3 $ \\
    \end{tabular}
\end{table}

where $k_1$ denotes the synthesis rate of $X_2$. $k_2$ is the degradation rate of $X_2$; $k_3$ is the rate of union of $X_1$ and $X_2$ to form $X_3$; $k_4$ is the separation of $X_3$ in $X_1$ and $X_2$; $k_5$ is the reaction rate of product $X_4$; and $k_6$ is the rate of union of $X_2$ and $X_4$ to obtain $X_3$.    

The stoichiometric coefficients are $\alpha_{ij}$ and $\beta_{ij}$, and the stoichiometric matrix are,

{\small
\begin{align}
 \alpha_{ij}&= \begin{pmatrix}
		0& 0 & 0 & 0  \\
            0& 1 & 0 & 0  \\
            1& 1 & 0 & 0  \\
            0& 0 & 1 & 0  \\
            0& 0 & 1 & 0  \\
            0& 1 & 0 & 1  \\
	\end{pmatrix} ,    &
 \beta_{ij}&=  \begin{pmatrix}
		0& 1 & 0 & 0  \\
            0& 0 & 0 & 0  \\
            0& 0 & 1 & 0  \\
            1& 1 & 0 & 0  \\
            0& 1 & 0 & 1  \\
            0& 0 & 1 & 0  \\  
	\end{pmatrix},  \nonumber \\
 \Gamma_{ij}&= \begin{pmatrix}
		  0& 0 & -1& 1 & 0 & 0 \\
		1& -1& -1& 1 & 1 &-1 \\
            0& 0 & 1 & -1& -1& 1 \\
            0& 0 & 0 & 0 & 1 &-1 \\
	\end{pmatrix}, & \nonumber
\end{align}}

the rates are 
\begin{align}
    R_1=& 1, & R_2&=  x_2 , \nonumber \\ 
    R_3=& x_1 x_2, & R_4&=  x_3 , \\
    R_5=&  x_3, & R_6&= x_2 x_4,  \nonumber 
\end{align}
where $x_1, x_2, x_3,x_4$ are the mean concentrations of $X_1,X_2,X_3,X_4$.

\subsubsection{Stationary State}
First, we analyze the stationary state using Proposition 5. The system satisfies the condition $\alpha_{ij}\leq 1$, but the variables are not independent because the range of the stoichiometric matrix is three, which is different from the number of chemical species, which is four. Now we reduced the number of chemical species because there is a conserved quantity $x_0= x_1 + x_3 + x_4$, then we substituted the value of  $x_4$ as $x_4= x_0- x_1 - x_3$ in the reaction rates, particularly only $R_6$ changes,
\begin{align}
    R'_6= x_2 (x_0- x_1 - x_3), 
\end{align}
and the stoichiometric matrix is reduced to
{\small
\begin{align}
 \Gamma'_{ij}&= \begin{pmatrix}
		  0& 0 & -1& 1 & 0 & 0 \\
		1& -1& -1& 1 & 1 &-1 \\
            0& 0 & 1 & -1& -1& 1 \\
	\end{pmatrix}, \nonumber
\end{align}}
with this expressions, we can now describe this system in the stationary state, and we get the following equations for mean concentrations
\begin{align}
    0=& -k_3 x_1 x_2 + k_4x_3, \nonumber \\
    0=& k_1- k_2x_2 -k_3 x_1 x_2 + k_4x_3 + k_5 x_3 - k_6 x_2 (x_0- x_1 - x_3), \nonumber \\
    0=& k_3 x_1 x_2 - k_4x_3 - k_5 x_3 + k_6 x_2 (x_0- x_1 - x_3) ,
\end{align}
and the next for the second central moments,

\begin{align}
    0=& \frac{1}{\Omega} (k_3 x_1 x_2 + k_4x_3) - 2 M^2_{1,1} k_3 x_2, \nonumber \\
    0=& \frac{1}{\Omega} (k_1 + k_2x_2 +k_3 x_1 x_2 + k_4x_3 + k_5 x_3 + k_6 x_2 (x_0- x_1 - x_3)) - 2 M^2_{2,2} (k_2 + k_3x_1 + k_6(x_0- x_1 - x_3)), \nonumber \\
    0=& \frac{1}{\Omega} (k_3 x_1 x_2 + k_4x_3 + k_5 x_3 + k_6 x_2 (x_0- x_1 - x_3)) - 2 M^2_{3,3} (k_4 + k_5 + k_6x_2).
\end{align}

Solving these equations, we get the following values for the stationary mean concentrations and the stationary second central moment, respectively 

\begin{align}
    x_{1,ss}=& \frac{k_4}{k_3} \frac{k_2}{k_1} x_{3,ss}, & M^2_{1,1,ss}=& \frac{x_{1,ss}}{\Omega}, \nonumber \\
    x_{2,ss}=& \frac{k_1}{k_2}, & M^2_{2,2,ss}=& \frac{x_{2,ss}}{\Omega}, \nonumber \\
    x_{3,ss}=& \frac{x_0}{1+  \frac{k_2}{k_1} \left( \frac{k_4 }{k_3}  +  \frac{k_5}{k_6}  \right) }  , & M^2_{3,3,ss}=& \frac{x_{3,ss}}{\Omega} \left( \frac{1}{1 + \frac{k_5}{k_4} + \frac{k_6}{k_4} \frac{k_1}{k_2}} \right).
\end{align}
All these results are exact in the stationary state. From these results, we can see that in the stationary state, the chemical species $x_1$ and $x_2$ follow a Poisson distribution, which is in agreement with the result of \cite{Grima}, in which, to obtain an exact description of the stationary state, it is indispensable to have a Poisson distribution for at least one of the variables.

\begin{figure*} [h!t]
  \begin{subfigure}{\linewidth}
\includegraphics[width=.33\textwidth]{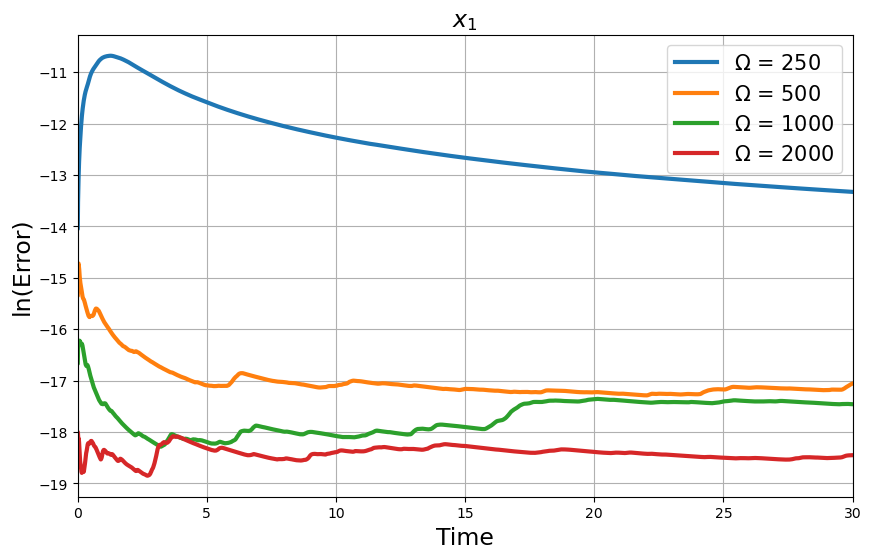}\hfill
\includegraphics[width=.33\textwidth]{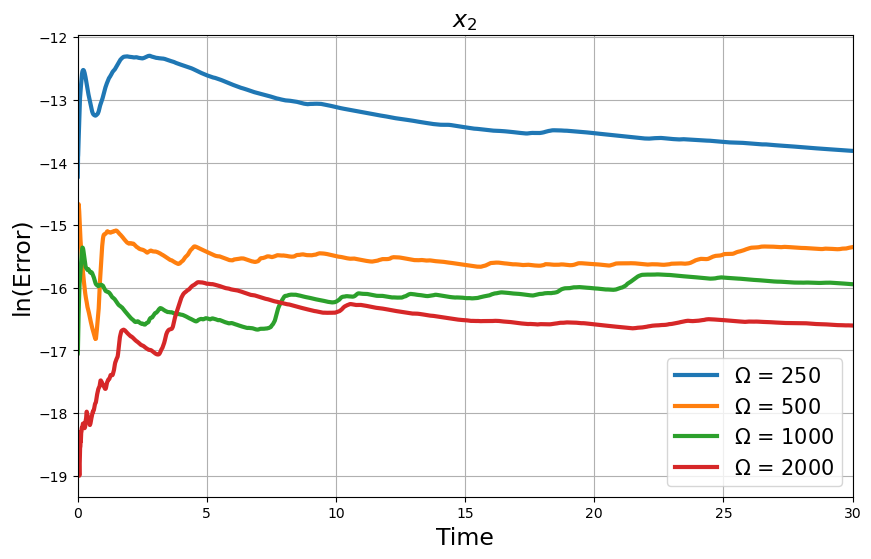}\hfill
\includegraphics[width=.33\textwidth]{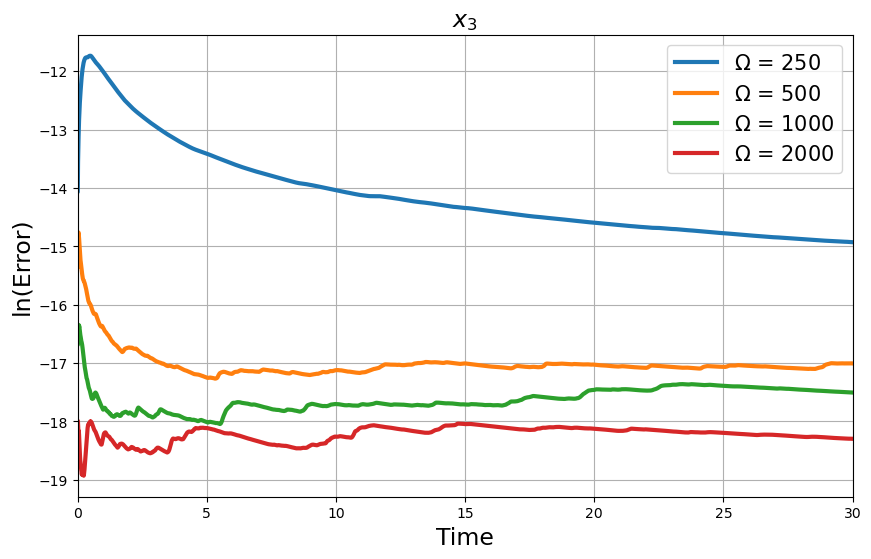}
\caption{\textbf{Mean concentration} }
  \end{subfigure}\par\medskip
    \begin{subfigure}{\linewidth}
\includegraphics[width=.33\textwidth]{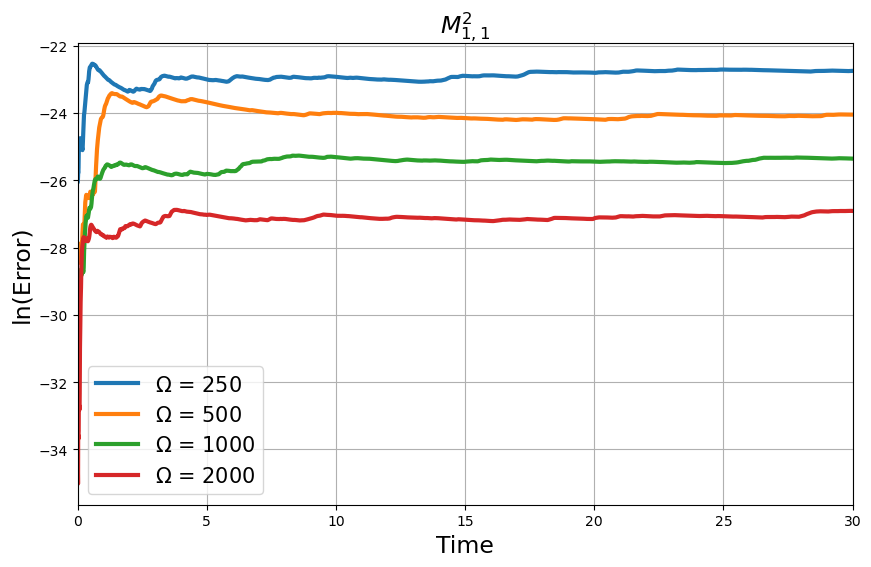}\hfill
\includegraphics[width=.33\textwidth]{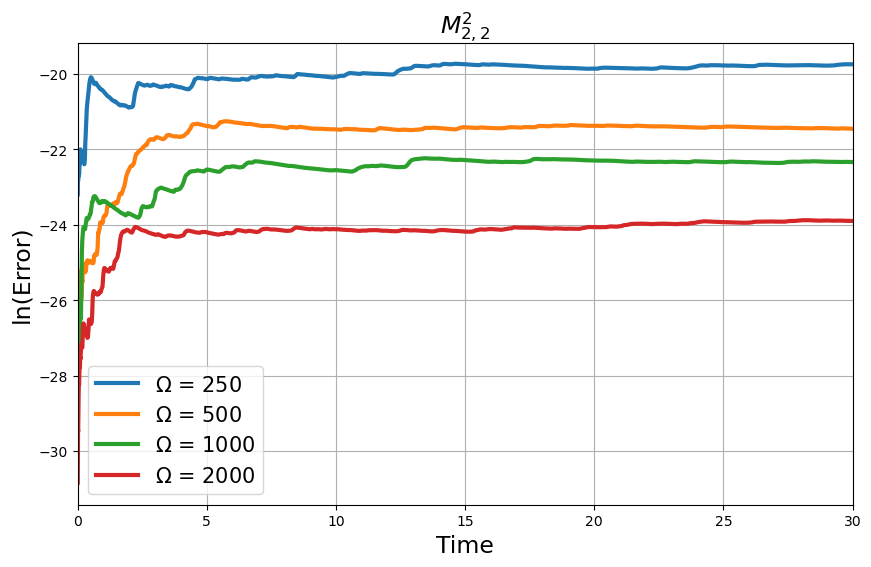}\hfill
\includegraphics[width=.33\textwidth]{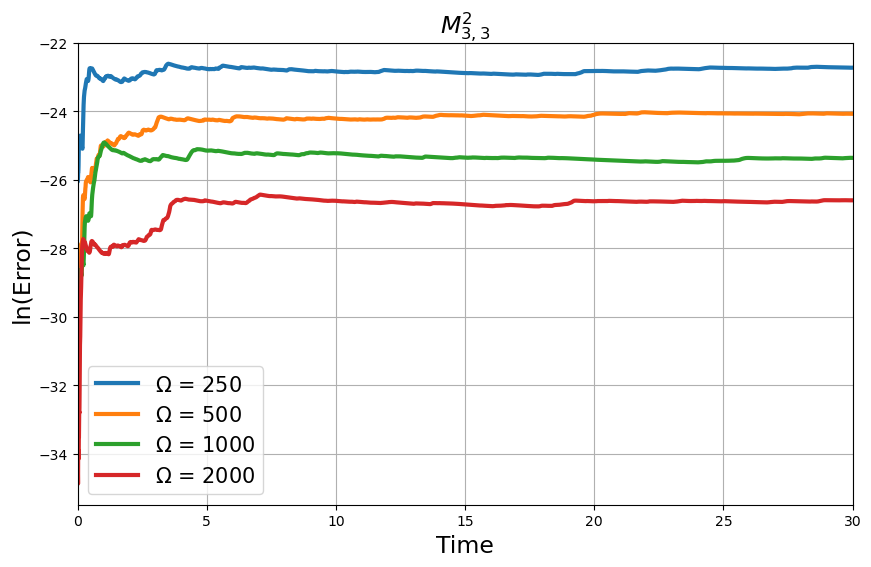}
\caption{\textbf{Second Central Moment} }
  \end{subfigure}
\caption{\textbf{Error in Enzymatic Processes.} In these figures, we graphed the error on a logarithmic scale at each instant in time, according to Equation (\ref{20}). In panel (a), we calculate the error of the mean of the variables, and in panel (b), we calculate the error of the variance of the variables.  We can see that when increases the size of the system $\Omega$, the error decreases. The parameters and initial conditions are listed in Table \ref{tabla1}. }
\label{fig.1}
\end{figure*}

\subsubsection{Dynamical Evolution}
We analyzed the dynamic evolution of the system using Equation (\ref{15}) until the second central moment, setting the diffusion parameters to zero. The error is then calculated using Equation (\ref{20}) by comparing the moment approach with the stochastic simulations.  

The results of this analysis are shown in Figure \ref{fig.1}. These figures show that the error everywhere is minimal. As the system size $\Omega$ increases, the error further decreases. Over time, the error diminishes but never reaches zero because of fluctuations that alter the dynamics and prevent the system from attaining a stationary state. Nevertheless, applying the second-moment approach to systems with second-order reactions provides a good approximation, because the error remains small.

\subsection{Antithetic} 

\begin{figure} [h!t]
\centering
\includegraphics[width=.3\textwidth]{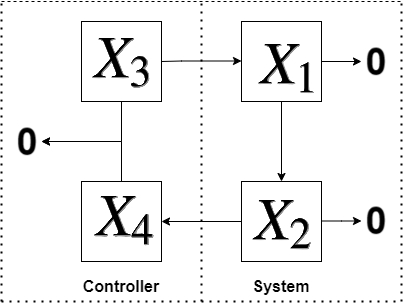}\hfill
\caption{\textbf{Antithetic.} In this figure, there is a representation of the antithetic, in which molecule $X_1$ produces $X_2$, which produces molecule $X_4$, $X_4$ binds $X_3$ to degrade, and $X_3$ produces molecule $X_1$. Negative feedback was observed in the part marked with the controller.  }
\label{fig.02}
\end{figure}

The second system we analyzed is the antithetic\cite{Briat}, which is presented as a synthetic controller that can be easily implemented in various systems, as shown in Figure \ref{fig.02}. Under the deterministic regime, this system exhibits either a stationary point for certain parameter regimes or a stable limit cycle for others. 

The system is described by the following reactions,

\begin{table}[h!]
    \centering
    \begin{tabular}{|c|c|}
        $ X_1 \stackbin{k_1}{\longrightarrow} X_1+X_2$ & $X_2  \stackbin{\gamma_p}{\longrightarrow} \emptyset$ \\
        $X_2   \stackbin{\theta_2}{\longrightarrow} X_2 + X_4 $ & $ \emptyset  \stackbin{\mu}{\longrightarrow} X_3 $ \\
        $X_3  \stackbin{\theta_1}{\longrightarrow} X_3+X_1$ & $ X_1 \stackbin{\gamma_p}{\longrightarrow} \emptyset $ \\
        $X_3 + X_4  \stackbin{\eta}{\longrightarrow} \emptyset $ &  \\
    \end{tabular}
\end{table}
where $k_1$ denotes the rate of synthesis of $X_2$ mediated by $X_1$, $\theta_2$ is the synthesis rate of $X_4$ mediate by $X_2$, $\theta_1$ is the synthesis rate of  $X_1$ mediate by $X_3$, $\eta$ is the degradation rate of the complex formed by $X_3$ and $X_4$, while $\gamma_p$ is the degradation rate of $X_1$ and $X_2$, and $\mu$ is the synthesis rate of $X_3$. 

The stoichiometric coefficients are $\alpha_{ij}$ and $\beta_{ij}$, and the stoichiometric matrix are,

{\small
\begin{align}
 \alpha_{ij}&= \begin{pmatrix}
		1& 0 & 0 & 0  \\
            0& 1 & 0 & 0  \\
            0& 1 & 0 & 0  \\
            0& 0 & 0 & 0  \\
            0& 0 & 1 & 0  \\
            1& 0 & 0 & 0  \\
            0& 0 & 1 & 1  \\
	\end{pmatrix} ,    &
 \beta_{ij}&=  \begin{pmatrix}
		1& 1 & 0 & 0  \\
            0& 0 & 0 & 0  \\
            0& 1 & 0 & 1  \\
            0& 0 & 1 & 0  \\
            1& 0 & 1 & 0  \\
            0& 0 & 0 & 0  \\
            0& 0 & 0 & 0  \\
	\end{pmatrix},  \nonumber \\
 \Gamma_{ij}&= \begin{pmatrix}
		0& 0 & 0 & 0 & 1 &-1 & 0   \\
            1&-1 & 0 & 0 & 0 & 0 & 0   \\
            0& 0 & 0 & 1 & 0 & 0 &-1  \\
            0& 0 & 1 & 0 & 0 & 0 &-1  
	\end{pmatrix}. & \nonumber
\end{align}}

The reaction rates of the system are,
\begin{align}
   R_1&=  x_1  , &
    R_2&= x_2, \nonumber \\
   R_3&=  x_2    , &
    R_4&= 1, \nonumber \\
    R_5&=  x_3   , &
    R_6&= x_1, \nonumber \\
    R_7&= x_3 x_4  , &
\end{align}
where $x_1, x_2, x_3,x_4$ are the mean concentrations of $X_1,X_2,X_3,X_4$.

\subsubsection{Stationary State}
First, we analyze the stationary state using Proposition 5. The system satisfied the condition $\alpha_{ij}\leq 1$, and the range of the stoichiometric matrix is four, the same number of different chemical species. Now we can describe this system in the stationary state, and we get the following equations for mean concentrations
\begin{align}
    0=& \theta_1 x_3 - \gamma_p x_1, \nonumber \\
    0=& k_1 x_1 - \gamma_p x_2, \nonumber \\
    0=& \mu  -\eta x_3 x_4, \nonumber \\
    0=& \theta_2 x_2  -\eta x_3 x_4,
\end{align}
and the next for the second central moment
\begin{align}
    0=& \frac{1}{\Omega}( \theta_1 x_3 + \gamma_p x_1) - 2M_{1,1} \gamma_p,   \nonumber \\
    0=&\frac{1}{\Omega}( k_1 x_1 + \gamma_p x_2) - 2M_{2,2} \gamma_p, \nonumber \\
    0=& \frac{1}{\Omega}(\mu   +\eta x_3 x_4) - 2M_{3,3} ( \eta x_4), \nonumber \\
    0=& \frac{1}{\Omega}(\theta_2 x_2  +\eta x_3 x_4) - 2M_{4,4} ( \eta  x_3),
\end{align}

solve these equations, we get the next values for the stationary mean concentrations and the stationary second central moments, respectively 
\begin{align}
    x_{1,ss}=& \frac{\gamma_p}{k_1}\frac{\mu}{\theta_2} , & M^2_{1,1,ss}=& \frac{x_{1,ss}}{\Omega}, \nonumber \\
    x_{2,ss}=&  \frac{\mu}{\theta_2}  , & M^2_{2,2,ss}=& \frac{x_{2,ss}}{\Omega}, \nonumber \\
     x_{3,ss}=& \frac{\gamma_p}{\theta_1}\frac{\gamma_p}{k_1}\frac{\mu}{\theta_2}  , & M^2_{3,3,ss}=& \frac{x_{3,ss}}{\Omega}, \nonumber \\
    x_{4,ss}=&  \frac{\theta_1}{\gamma_p}\frac{k_1}{\gamma_p}\frac{\theta_2}{\eta}   , & M^2_{4,4,ss}=& \frac{x_{4,ss}}{\Omega},
\end{align}

these equations are the exact mean concentrations and second central moment in the stationary state.  Similar to the case in the previous section, the chemical species follow a Poisson distribution, which is in concordance with the result in \cite{Grima}, in which, to obtain an exact description of the stationary state, it is indispensable to have a Poisson distribution for at least one variable.

\begin{figure*} [h!t]
\begin{subfigure}{\linewidth}
\includegraphics[width=.24\textwidth]{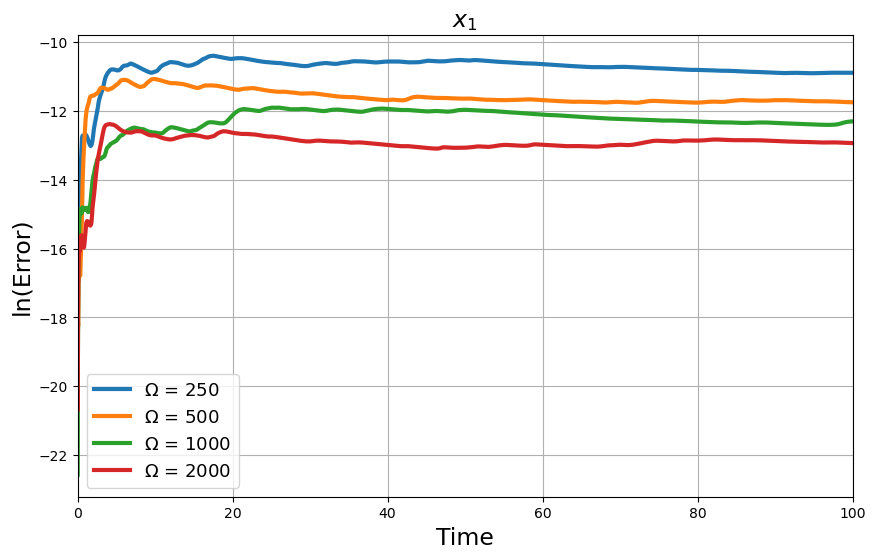}\hfill
\includegraphics[width=.24\textwidth]{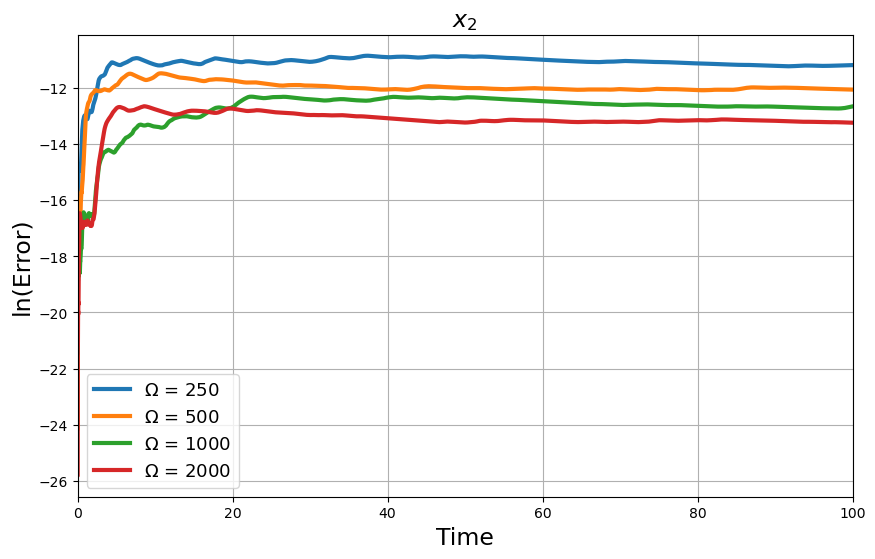}\hfill
\includegraphics[width=.24\textwidth]{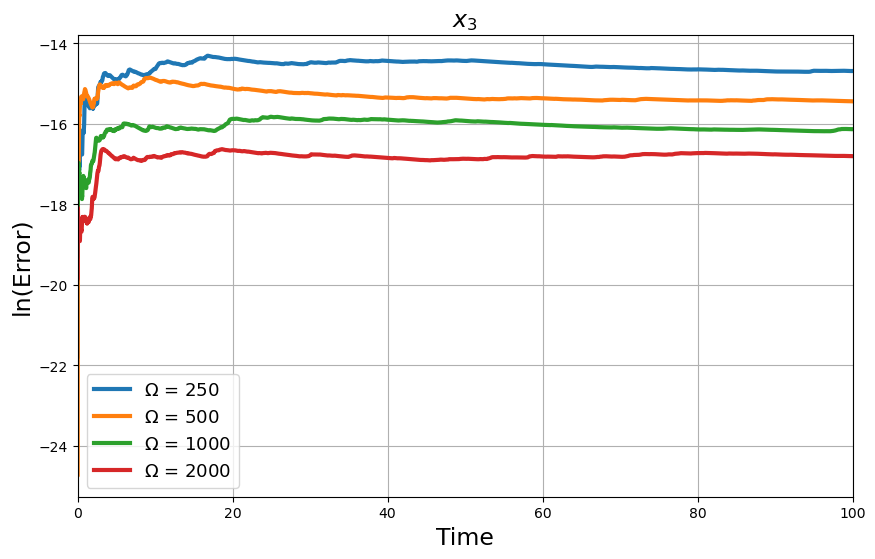}\hfill
\includegraphics[width=.24\textwidth]{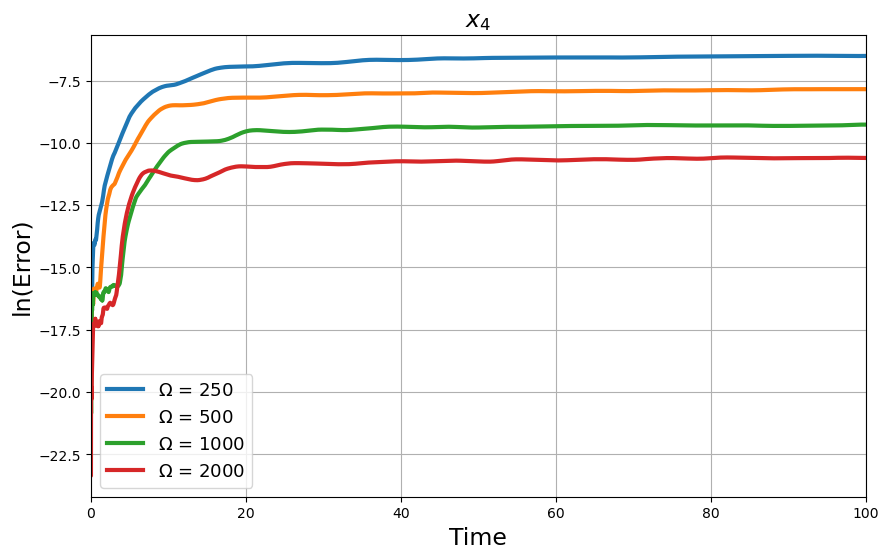}
\caption{\textbf{Mean concentration} }
  \end{subfigure}\par\medskip
    \begin{subfigure}{\linewidth}
\includegraphics[width=.24\textwidth]{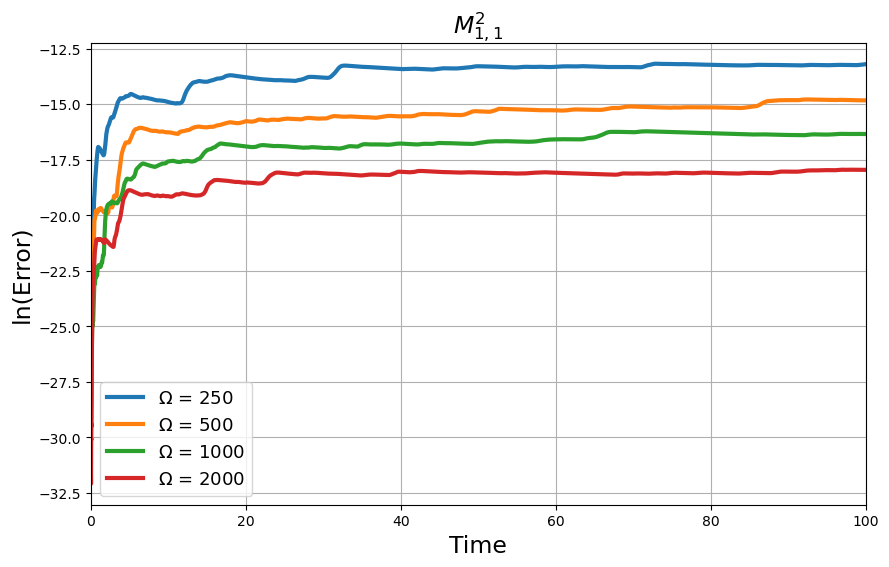}\hfill
\includegraphics[width=.24\textwidth]{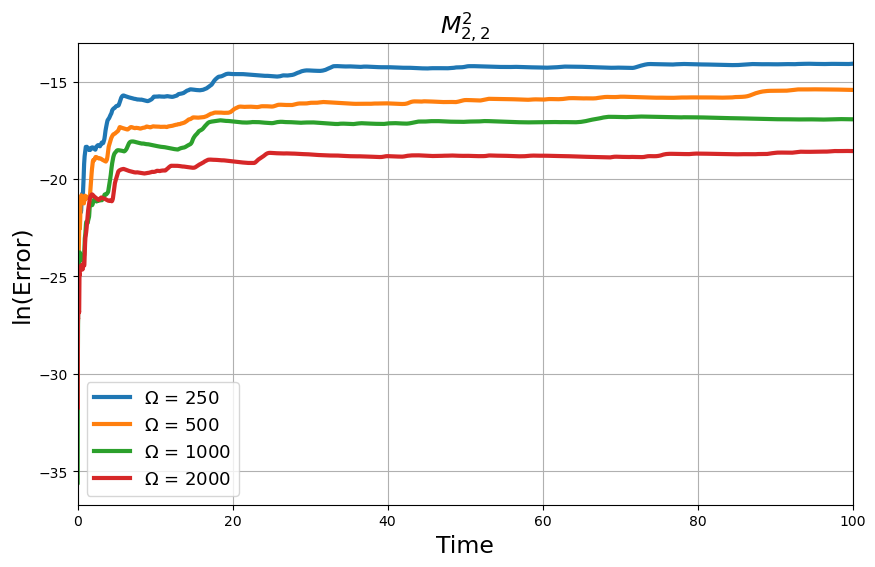}\hfill
\includegraphics[width=.24\textwidth]{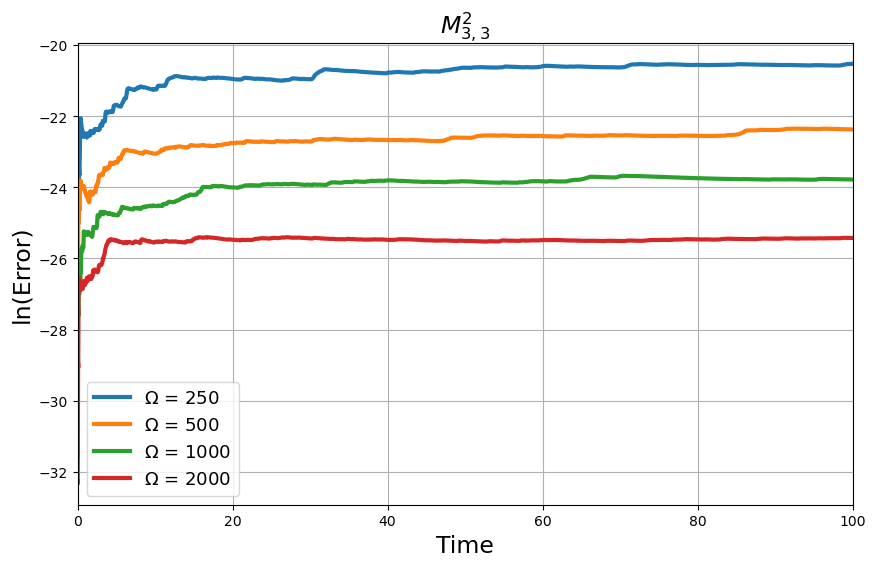}\hfill
\includegraphics[width=.24\textwidth]{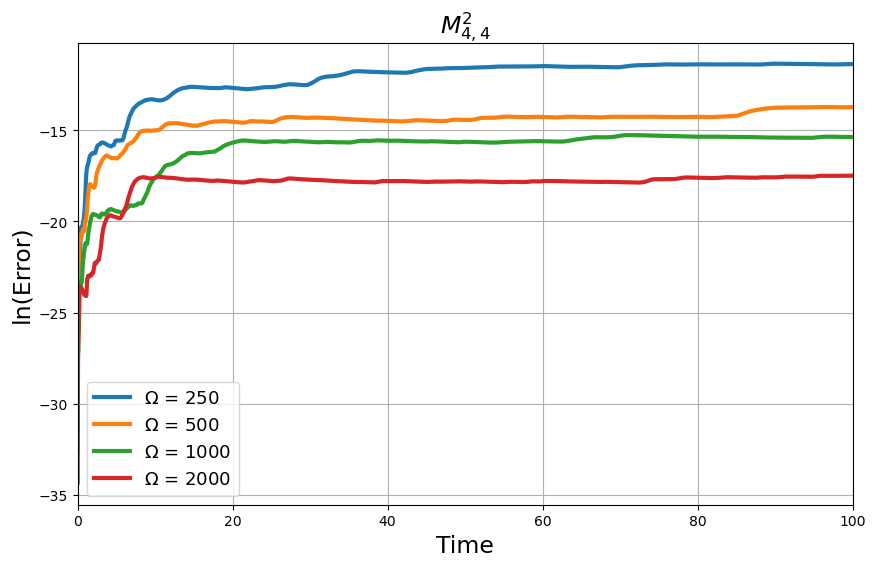}
\caption{\textbf{Second Central Moment} }
  \end{subfigure}
\caption{\textbf{Error in Antithetic.} In these figures, we graphed the error on a logarithmic scale at each instant in time according to Equation (\ref{20}). In panel (a), we calculate the error of the mean of the variables, and in panel (b),  we calculate the error of the variance of the variables.  We can see that when increases the size of the system $\Omega$, the error decreases. The parameters and initial conditions are listed in Table \ref{tabla2}.}
\label{fig.2}
\end{figure*}

\subsubsection{Dynamical Evolution}
We analyzed the dynamic evolution of the system using Equations (\ref{13}) and (\ref{14}), setting the diffusion parameters to zero. The error is then calculated using Equation (\ref{20}) by comparing the moment approach with the stochastic simulations. This system has two types of behavior: one of which has a stable equilibrium point, and another in which there is a stable limit cycle. Then, we analyze the behavior of the system in these two scenarios; the last is in Appendix \ref{F}.  The results of analyzing the system when it reaches a stationary point are shown in Figure \ref{fig.2}. The error at each time point was very small. As the system size $\Omega$ increases, the error decreases further. Over time, the error also diminishes, but never reaches zero, owing to fluctuations that alter the dynamics.     

By comparing Figures \ref{fig.2} and \ref{fig.f1}, we observe that when the parameters result in a stable equilibrium point, the system is accurately described by increasing $\Omega$, further improving precision. However, when the parameters lead to a stable limit cycle, the accuracy decreases, particularly in terms of variance, which increases with mean concentration. Further investigation is required to address this issue.

In the examples analyzed, we start with Proposition 5. However, alternative methods exist for determining stationary distributions from which the means and central moments can be derived. One such approach is based on the work of Kurtz et al. \cite{Anderson}, which relies on systems with zero deficiency and weak reversibility. These properties are satisfied by the examples presented in this work, suggesting that future studies could explore the connection between these two methodologies.

\subsection{Genetic Regulatory network with Hill functions}

\begin{figure} [h!t]
\centering
\includegraphics[width=.25\textwidth]{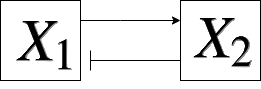}\hfill
\caption{\textbf{Negative Feedback.}  In this figure, there is a representation of a genetic regulatory network with negative feedback, where $X_1$ suppresses the production of $X_2$ and $X_2$ promotes the production of  $X_1$.}
\label{fig.03}
\end{figure}

In the previous sections, we analyzed systems with only mass-action propensity functions, and we now provide the conditions for obtaining an exact approach for systems that involve Hill functions, which is important because this type is common in regulatory genetic networks. We analyzed a system with negative feedback, as shown in Figure \ref{fig.03}. Negative feedback in genetic regulatory networks confers evolutionary resilience, noise control, and functional versatility, making it a fundamental motif in biological systems \cite{Marciano, Hinczewski} and can present oscillations \cite{Kulasiri}.  \\

We consider a simplified system in which we consider only the dynamics of the proteins. Then, the system is described by the following reactions,

\begin{table}[h!]
    \centering
    \begin{tabular}{|c|c|}
        $ \emptyset \stackbin{k_1}{\longrightarrow} X_1$ & $X_1  \stackbin{k_2}{\longrightarrow} \emptyset$ \\
        $0   \stackbin{k_3}{\longrightarrow} X_2  $ & $ X_2  \stackbin{k_4}{\longrightarrow} \emptyset $ \\
    \end{tabular}
\end{table}

the stoichiometric coefficients are $\alpha_{ij}$ and $\beta_{ij}$, and the stoichiometric matrix are,

{\small
\begin{align}
 \alpha_{ij}&= \begin{pmatrix}
		0& 0   \\
            1& 0   \\
            0& 0   \\
            0& 1   \\
	\end{pmatrix} ,    &
 \beta_{ij}&=  \begin{pmatrix}
		1& 0   \\
            0& 0   \\
            0& 1   \\
            0& 0   \\
	\end{pmatrix},  &
 \Gamma_{ij}&= \begin{pmatrix}
		1& -1 & 0 & 0   \\
            0 & 0 & 1& -1      \\
	\end{pmatrix}. \nonumber
\end{align}}

The reaction rates of the system,
\begin{align}
   R_1&= 1, &
    R_2&=  x_1, \nonumber \\
   R_3&= 1, &
    R_4&=  x_2, 
\end{align}
where  $x_1, x_2$ are the mean concentrations of $X_1, X_2$ and the parameters $k_1$ and $k_3$  are functional rates that have the following form,

{\small
\begin{align}
    k_1= k_1^*  \left( \frac{1}{1+ (x_2^3 + 3x_2 M^2_{2,2} + M^3_{2,2,2}) -  \frac{3}{\Omega}(x_2^2 + M^2_{2,2} ) + \frac{2}{\Omega^2}x_2 } \right),    \nonumber \\
    k_3= k_3^* \left( \frac{ (x_1^3 + 3x_1 M^2_{1,1} + M^3_{1,1,1}) - \frac{3}{\Omega}(x_1^2 + M^2_{1,1} ) + \frac{12}{\Omega^2}x_1}{1+ (x_1^3 + 3x_1 M^2_{1,1} + M^3_{1,1,1}) -  \frac{3}{\Omega}(x_1^2 + M^2_{1,1} ) + \frac{2}{\Omega^2}x_1 } \right).
\end{align}}
Where $x_1$, $M^2_{1,1}$, $ M^3_{1,1,1}$ are the mean concentration, second central moment, and third central moment of $X_1$, respectively, some similar for $X_2$. The terms in parentheses are Hill functions; for more details on their derivations, refer to Appendix \ref{C}. We chose a coefficient Hill $n=3$ for both. The first Hill function is for a repressor and the next function is for an activator. These Hill functions are exact because they account for all terms without any approximation, including the third central moment.

\subsubsection{Exact approach} 
This system satisfies the conditions of Proposition 1 and can be described exactly up to the third central moment. By setting the diffusion parameters to zero, the equations for the mean concentrations are as 
\begin{align}
    \frac{\partial x_1}{ \partial t}= k_1 - k_2 x_1 ,\nonumber \\
    \frac{\partial x_2}{ \partial t}= k_3  - k_4 x_2, \label{30}
\end{align}
for the second central moment
\begin{align}
        \frac{\partial M^2_{1,1}}{ \partial t}= \frac{1}{\Omega} \left( k_1 + k_2 x_1 \right) - 2 k_2 M^2_{1,1},\nonumber \\
    \frac{\partial M^2_{2,2}}{ \partial t}= \frac{1}{\Omega} \left( k_3 +k_4 x_2 \right) - 2 k_4 M^2_{2,2}, \label{31}
\end{align}
and for the third central moment
\begin{align}
    \frac{\partial M^3_{1,1,1}}{ \partial t}=&\frac{1}{\Omega^2} \left( k_1 - k_2 x_1 \right) + \frac{3}{\Omega} k_2 M^2_{1,1} \nonumber \\
    &+ \frac{3M^2_{1,1}}{\Omega} \left( k_1 - k_2 x_1\right)  - 3k_2M^3_{1,1,1},\nonumber \\ 
    \frac{\partial M^3_{2,2,2}}{ \partial t}=& \frac{1}{\Omega^2} \left( k_3  -k_4 x_2 \right)  + \frac{3}{\Omega} k_4 M^2_{2,2} \nonumber \\
    &+ \frac{3 M^2_{2,2}}{\Omega}\left( k_3-k_4 x_2 \right) - 3 k_4 M^3_{2,2,2}, \label{32} 
\end{align}

Equations (\ref{30})–(\ref{32}) are exact and provide an accurate description of the system dynamics. This analysis focuses on a specific case, in which the Hill function explicitly depends on the third central moment. However, this framework can be extended to account for higher-order central moments if the Hill function is dependent on these moments.

\begin{figure*} [h!t]
\begin{subfigure}{\linewidth}
\includegraphics[width=.33\textwidth]{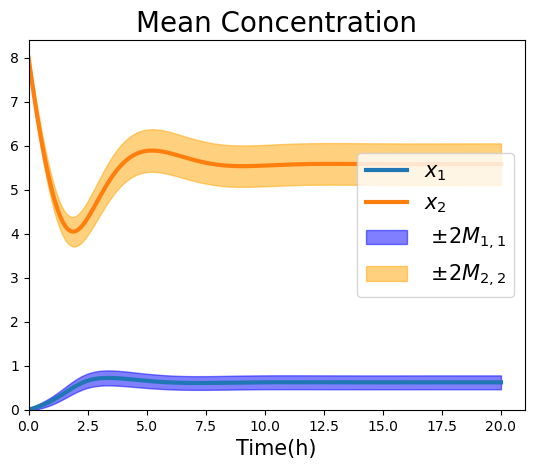}\hfill
\includegraphics[width=.33\textwidth]{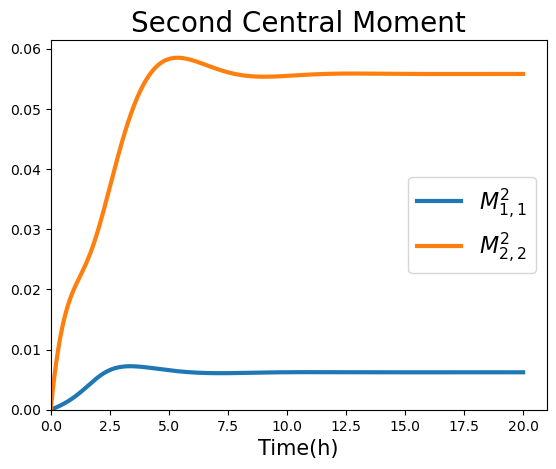}\hfill
\includegraphics[width=.33\textwidth]{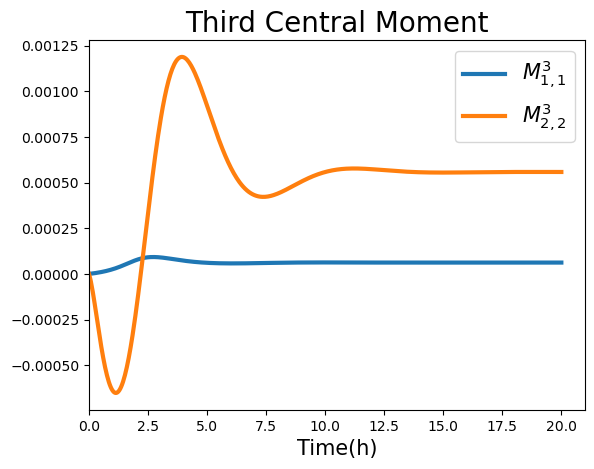}
\end{subfigure}
\caption{\textbf{Negative Feedback.} These figures show the results of the numerical solution of equations (\ref{30})–(\ref{32}), depicting the mean concentration, second central moment, and third central moment. It is important to note that the dynamics are exact within the framework presented in this study. The parameters and initial conditions used for the simulations are provided in Table \ref{tabla3}.}
\label{fig.4}
\end{figure*}

From Equations (\ref{30})-(\ref{32}) we can analyze the stationary state, then we get for the central moments,
\begin{align}
    M^2_{1,1,ss}=& \frac{x_{1,ss}}{\Omega}, & M^3_{1,1,1,ss}=& \frac{x_{1,ss}}{\Omega^2}, \nonumber \\
    M^2_{2,2,ss}=& \frac{x_{2,ss}}{\Omega}, & M^3_{2,2,2,ss}=& \frac{x_{2,ss}}{\Omega^2}.
\end{align}
From these results, we can conclude that each variable has a Poisson distribution in the stationary state. To obtain the values of $x_{1,ss}$ and $x_{2,ss}$ we need to solve Equation (\ref{30}) set equal to zero, and substituting the values of the central moments, we obtain the following equations:
\begin{align}
        0= k_1^*\frac{1}{1+x_{2,ss}^3} - k_2 x_{1,ss}, \nonumber \\
        0= k_3^*\frac{x_{1,ss}^3}{1+x_{1,ss}^3}  - k_4 x_{2,ss}. 
\end{align}
These equations are the same as those used in the deterministic description. From this result, we can conclude that moment-based analysis provides the same equations as the deterministic part for the stationary state, where fluctuations do not affect the stationary points. Using Equations (\ref{30})–(\ref{32}), we numerically solved the dynamics of the system for each variable. The results are shown in Figure \ref{fig.4}, the mean concentration and fluctuations are in the shaded area. It is noticeable that the error does not increase over time and remains limited, unlike other schemes that employ closure techniques \cite{Lakatos}.

\section{Genetic Regulatory Network with Hill Functions and Diffusion}  \label{section6}
In the previous sections, we analyzed only systems without diffusion, but provided the conditions for obtaining an exact approach for systems that involve Hill functions in Proposition 1. Even if the system includes diffusion, Hill functions are important because this type is common in regulatory genetic networks. Next, we analyzed a genetic regulatory network with negative feedback through diffusion. Negative feedback in genetic regulatory networks confers evolutionary resilience, noise control, and functional versatility, and presents oscillations, making them a fundamental motif in biological systems \cite{Marciano, Hinczewski}. However, by including the spatial domain, these systems exhibit richer dynamics, including pattern formation \cite{Turing}.  By considering the spatial domain, we aim to better understand how local regulatory interactions and diffusive coupling shape the global behavior of genetic networks, ultimately providing insights into organismal development and the spatial organization of cellular functions \cite{Jong}.

Now, we analyze a genetic regulatory network with negative feedback, similar to the previous model, where we only considered the dynamics of the proteins, where the parameters $k_1^r$ and $k_3^r$  are functional parameters that have Hill-type functions, we chose a Hill coefficient $n=2$, then we have
\begin{align}
    k_1^{r}=& k_1^* \left( \frac{1}{1+ ((x_2^r)^2 +  M^2_{2^r,2^r} ) - \frac{1}{\Omega}(x_2^r ) } \right),    \nonumber \\
    k_3^{r}=& k_3^* \left( \frac{ ((x_1^r)^2 +  M^2_{1,1} ) - \frac{1}{\Omega}(x_1^r)}{1+ ((x_1^r)^2 +  M^2_{1^r,1^r} ) - \frac{1}{\Omega}(x_1^r) } \right) ,\label{38}
\end{align}
where $x_1^r$ is the mean concentration of $X_1$ in voxel $r$ and $M^2_{1^r,1^r}$ is the second central moment of $X_1$ in voxel $r$ and similarly for $x_2^r$ and $M^2_{2^r,2^r}$. The terms in parentheses are Hill functions; for more details on their derivations, refer to Appendix \ref{C}. The first Hill function is for a repressor, and the next function is for an activator. These Hill functions are exact because they account for all terms without any approximation, including the second central moment.  The only difference between the genetic regulatory network with negative feedback from the previous section is that effective parameters depend only on the second central moment.

\begin{figure*} [h!t]
\centering
\includegraphics[width=.8\textwidth]{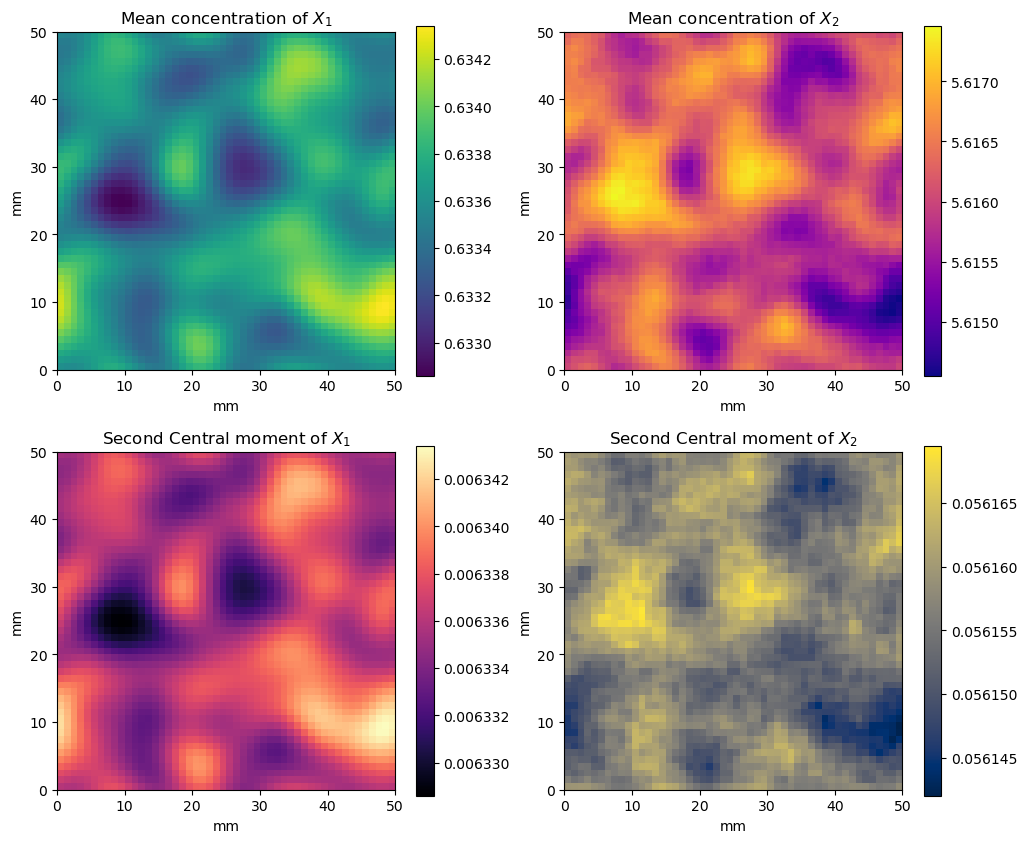}
\caption{\textbf{Negative Feedback with diffusion.}  In these figures, we obtain the results of the numerical solutions of equations (\ref{38})–(\ref{41}). We observed that closer to the edge, the average concentrations, as well as the second central moment, have different values with respect to the center.  The parameters and initial conditions used are listed in Table \ref{tabla4}.}
\label{fig.5}
\end{figure*}

\subsection{Exact Approach}
To describe the system, we need a differential equation up to the second central moment; in this case, the equations for the mean concentration are
\begin{align}
    \frac{\partial x_1^r }{ \partial t}= k_1^{r} - k_2 x_1^r + \sum_q ( d^1_{qr} x_1^q - d^1_{rq} x_1^r), \nonumber \\
    \frac{\partial x_2^r }{ \partial t}= k_3^{r} - k_4 x_2^r + \sum_q ( d^2_{qr} x_2^q - d^2_{rq} x_2^r). 
\end{align}
And for the second central moment,

\begin{align}
    \frac{\partial M^2_{1^{r_1},1^{r_2}}}{ \partial t}=& \frac{\delta_{r_1,r_2}}{\Omega} \left( k_1^{r} + k_2 x_1^{r_1} \right) - 2 k_2 M^2_{1^{r_1},1^{r_2}} + \frac{1}{\Omega} \left( \delta_{r_1,r_2} \sum_q ( d^1_{qr_1} x_1^q + d^1_{r_1q} x_1^{r_1}) - (d^1_{r_1 r_2} x_1^{r_1} + d^1_{r_2 r_1} x_1^{r_2})  \right) \nonumber \\
    +& \sum_q ( d^1_{qr_1} M^2_{1^{q},1^{r_2}} - d^1_{r_1q} M^2_{1^{r_1},1^{r_2}} ) + \sum_q ( d^1_{qr_2} M^2_{1^{r_1},1^{q}} - d^1_{r_2q} M^2_{1^{r_1},1^{r_2}} ), \nonumber \\
    \frac{\partial M^2_{2^{r_1},2^{r_2}}}{ \partial t}=& \frac{\delta_{r_1,r_2}}{\Omega} \left( k_3^{r} + k_4 x_1^{r_1} \right) - 2 k_4 M^2_{2^{r_1},2^{r_2}} + \frac{1}{\Omega} \left( \delta_{r_1,r_2} \sum_q ( d^2_{qr_1} x_2^q + d^2_{r_1q} x_2^{r_1}) - (d^2_{r_1 r_2} x_2^{r_1} + d^2_{r_2 r_1} x_2^{r_2})  \right) \nonumber \\
    +& \sum_q ( d^2_{qr_1} M^2_{2^{q},2^{r_2}} - d^2_{r_1q} M^2_{2^{r_1},2^{r_2}} ) + \sum_q ( d^2_{qr_2} M^2_{2^{r_1},2^{q}} - d^2_{r_2q} M^2_{2^{r_1},2^{r_2}} ). \label{41}
\end{align}

The second central moments are calculated between $X_1$ in voxels $r_1$ and $r_2$. We did not include other second central moments because these do not affect the dynamics of the mean concentrations. Equations (\ref{41}) are exact, allowing for the quantification of fluctuations during the dynamical evolution. Their form is particularly suitable for numerical solutions. To simplify the computation, we assume that the species only move between first neighbors, that is, 
\begin{align}
    d^i_{r_1,r_2}= 
\begin{cases}         
\frac{D_i}{\lambda} & \text{if } |r_1-r_2|=1  \\    
0 & \text{other cases}    
\end{cases}  .
\end{align}
where $i=1,2$, $D_i$ the parameter of diffusion and $\lambda$ is the size of each voxel. 

Figure \ref{fig.5} shows the numerical solutions of Equations (\ref{38})–(\ref{41}), illustrating the mean concentration of each variable and its second central moment in each voxel. These results correspond near to the stationary state. Notably, there are some variations in the mean concentrations and second central moment because the system does not reach the stationary state. However, the values of the second central moments and the mean concentration had a proportion of 1:100. 

\section{Conclusions} \label{section7}

In this study, we established specific conditions under which a chemical reaction-diffusion network can be described exactly up to the second central moment, avoiding the problem of moment closure. We explored these conditions in two scenarios. First, when the kinetic parameters deviate from the mass-action kinetics, then there are effective parameters dependent on the means and higher-order moments. The second scenario is when a system follows mass-action kinetics.  For the first case, Proposition 1 outlines the condition for exactness when higher-order moments are involved; this type of system is common and applicable to systems such as genetic regulatory networks. Proposition 5 provides criteria for exact descriptions of stationary states, even in systems with higher-order reactions.

The analysis of the antithetic system under the conditions of Proposition 5 reveals that the stationary values of the mean concentrations and second central moments can be computed exactly, yielding a Poisson-distributed species. This result provides an example of the exact analytical characterization of biochemical networks with feedback. However, the emergence of a stable limit cycle under different parameter regimes highlights the limitations of the current framework in describing non-stationary dynamics.
This limitation highlights the need for further investigations to identify the underlying causes and potential solutions for these systems.

The application of Proposition 1 to a genetic regulatory network with negative feedback shows that the dynamics of the system can be exactly described, even in the presence of nonlinear regulation through Hill functions and diffusion. By systematically linking the degree of moment closure to the Hill function's dependence, this framework allows for an exact characterization of the stochastic behavior of the system while capturing intrinsic fluctuations. This result provides a generalizable analytical tool for modeling regulatory networks with or without spatial effects, enabling rigorous stability analysis in systems with fluctuations.

The propositions and their corollaries presented here are significant contributions to the field, as they provide a robust framework for describing a class of systems with exact mean and second central moments. This approach not only helps quantify fluctuations but also enables stability analyses similar to those in deterministic systems. Because the system is now a set of ordinary differential equations, solving it is computationally more efficient. These insights can pave the way for extending this methodology to more complex systems and broader applications in diverse scenarios. However, it is important to mention that this approach cannot be applied to some systems, for example, systems with bimodality, and further investigation is necessary.

\section*{Acknowledgments}
Manuel E. Hernández-García acknowledges the financial support of SECITHI through the program "Becas Nacionales 2023".

Jorge Velázquez-Castro acknowledges financial support of BUAP-VIEP through project 00398-PV.


\section*{Declarations}
The authors declare no conflicts of interest regarding the publication of this article. 

All data generated or analyzed in this study are included in this published article.

\appendix

\section{Prof of Proposition 1 \label{A}}
Before presenting the proof, we provide a formula for describing the dynamic evolution of the $m$-th moment. To achieve this, we use the master equation and define \(\eta_j^r = (S_{j}^r - \braket{S_{j}^r})\), leading to the following expression only for reactions:  

{\tiny
\begin{align}
    &\frac{\partial M^m_{j_1^{r_1}, j_2^{r_2}, \dots , j_m^{r_m}}}{\partial t}= \frac{\Omega}{\Omega^m} \sum_i \sum_r \langle \left( (\eta_{j_1}^{r_1} + \Gamma_{j_1i}^{r_1} ) (\eta_{j_2}^{r_2} + \Gamma_{j_2i}^{r_2} ) \dots (\eta_{j_m}^{r_m} + \Gamma_{j_m i}^{r_m} ) - (\eta_{j_1}^{r_1} \eta_{j_2}^{r_2} \dots \eta_{j_m}^{r_m} )  \right) a_i^r(\mathbf{S})   \rangle \nonumber \\
    =& \frac{\Omega}{\Omega^m} J^m \left(\sum_i \sum_r \langle \left( (\eta_{j_1}^{r_1} \eta_{j_2}^{r_2} \dots \eta_{j_{m-1}}^{r_{m-1}} ) \Gamma_{j_mi}^{r_m} \delta_{r_m,r}  +  (\eta_{j_1}^{r_1} \eta_{j_2}^{r_2} \dots \eta_{j_{m-2}}^{r_{m-2}} ) \Gamma_{j_{m-1}i}^{r_m} \Gamma_{j_mi}^{r_m} \delta_{r_{m-1},r_m,r}  +...+ \frac{\delta_{r_1, r_2, \dots,r_m,r}}{m} \Gamma_{j_1i}^{r_1} \Gamma_{j_2i}^{r_1} \dots \Gamma_{j_mi}^{r_1} \right) a_i^r(\mathbf{S})   \rangle \right), \label{a.1}
\end{align}}

where $J^m$ represents the permutation of all $m$ free indices. If all of the reactions that have the system are until order one, then the propensity rates are 
\begin{align}
    a_i^r(\mathbf{S})= k_i^{r} \prod_j \left( \frac{S_j^r}{\Omega} \right)^{\alpha_{ij}}, 
\end{align}
from which we obtain the deterministic rates exactly as
\begin{align}     
R_i^r(\mathbf{s})= \frac{a_i^r(\braket{\mathbf{S}})}{k_i^{r}}=  \prod_{j=1}^{N} (s_j^r)^{\alpha_{ij}}, 
\end{align} 
then, making a Taylor expansion of it around the mean, we get

{\footnotesize
\begin{align}
    a_i^r(\mathbf{S}^r)= a_i^r(\braket{\mathbf{S}^r}) + \sum_j \eta_j^r \frac{\partial a_i^r(\braket{\mathbf{S}^r})}{\partial S_j^r} = k_i^{r} \left( R_i^r(\mathbf{s}^r) + \sum_j  \frac{\eta_j^r}{\Omega} \frac{\partial R_i^r(\mathbf{s}^r)}{\partial s_j^r} \right),
\end{align}}
note, that this expansion is exact because $R_i^r(\mathbf{s})$ is linear. Now we use this result in Equation (\ref{a.1}), then we get

{\tiny
\begin{align}
   & \frac{\partial M^m_{j_1^{r_1}, j_2^{r_2}, \dots , j_m^{r_m}}}{\partial t}= \frac{\Omega}{\Omega^m} J^m \left(\sum_i \sum_r \langle \left( (\eta_{j_1}^{r_1} \eta_{j_2}^{r_2} \dots \eta_{j_{m-1}}^{r_{m-1}} ) \Gamma_{j_mi}^{r_m} \delta_{r_m,r}  +   (\eta_{j_1}^{r_1} \eta_{j_2}^{r_2} \dots \eta_{j_{m-2}}^{r_{m-2}} ) \Gamma_{j_{m-1}i}^{r_m} \Gamma_{j_mi}^{r_m} \delta_{r_{m-1},r_m,r}  +...+ \frac{\delta_{r_1, r_2, \dots,r_m,r}}{m} \Gamma_{j_1i}^{r_1} \Gamma_{j_2i}^{r_1} \dots \Gamma_{j_mi}^{r_1} \right) a_i^r(\mathbf{S})   \rangle \right)\nonumber \\
    =& \frac{\Omega}{\Omega^m} J^m \left(\sum_i \sum_r \langle \left( (\eta_{j_1}^{r_1} \eta_{j_2}^{r_2} \dots \eta_{j_{m-1}}^{r_{m-1}} ) \Gamma_{j_mi}^{r_m}\delta_{r_m,r}  +  (\eta_{j_1}^{r_1} \eta_{j_2}^{r_2} \dots \eta_{j_{m-2}}^{r_{m-2}} ) \Gamma_{j_{m-1}i}^{r_m} \Gamma_{j_mi}^{r_m} \delta_{r_{m-1},r_m,r} + \dots + \frac{\delta_{r_1, r_2, \dots,r_m,r}}{m} \Gamma_{j_1i}^{r_1} \Gamma_{j_2i}^{r_1} \dots \Gamma_{j_mi}^{r_1} \right) k_i^{r} \left( R_i^r(\mathbf{s})  + \sum_j  \frac{\eta_j^r}{\Omega} \frac{\partial R_i^r(\mathbf{s})}{\partial s_j^r} \right)   \rangle  \right) \nonumber \\
    =& \frac{\Omega}{\Omega^m} J^m \left(\sum_i \sum_r \langle \left( (\eta_{j_1}^{r_1} \eta_{j_2}^{r_2} \dots \eta_{j_{m-1}}^{r_{m-1}} ) \Gamma_{j_mi}^{r_m}\delta_{r_m,r}  +  (\eta_{j_1}^{r_1} \eta_{j_2}^{r_2} \dots \eta_{j_{m-2}}^{r_{m-2}} ) \Gamma_{j_{m-1}i}^{r_m} \Gamma_{j_mi}^{r_m} \delta_{r_{m-1},r_m,r} + \dots + \frac{\delta_{r_1, r_2, \dots,r_m,r}}{m} \Gamma_{j_1i}^{r_1} \Gamma_{j_2i}^{r_1} \dots \Gamma_{j_mi}^{r_1} \right) \left( k_i^{r} R_i^r(\mathbf{s}) \right)   \rangle  \right) \nonumber \\ 
    +& \frac{1}{\Omega^m} J^m \left(\sum_i \sum_j \sum_r \langle \left( (\eta_{j_1}^{r_1} \eta_{j_2}^{r_2} \dots \eta_{j_{m-1}}^{r_{m-1}}\eta_j^r ) \Gamma_{j_mi}^{r_m}\delta_{r_m,r} +  (\eta_{j_1}^{r_1} \eta_{j_2}^{r_2} \dots \eta_{j_{m-2}}^{r_{m-2}} \eta_j^r ) \Gamma_{j_{m-1}i}^{r_m} \Gamma_{j_mi}^{r_m}\delta_{r_{m-1},r_m,r} + \dots + \frac{\delta_{r_1, r_2, \dots,r_m,r}}{m} \Gamma_{j_1i}^{r_1} \Gamma_{j_2i}^{r_1} \dots \Gamma_{j_mi}^{r_1} \eta_j^r \right) \left( k_i^{r}  \frac{\partial R_i^r(\mathbf{s})}{\partial s_j^r} \right)   \rangle  \right) \nonumber \\ 
    =&    J^m \left(\sum_i \sum_r \left( M^{m-1}_{j_1^{r_1}, j_2^{r_2}, \dots j_{m-1}^{r_{m-1}}} {\Gamma_{j_mi}^{r_m}\delta_{r_m,r}}  +  M^{m-2}_{j_1^{r_1}, j_2^{r_2}, \dots j_{m-2}^{r_{m-2}}} \frac{\Gamma_{j_{m-1}i}^{r_m} \Gamma_{j_mi}^{r_m} \delta_{r_{m-1},r_m,r}}{\Omega}  + \dots + \frac{\delta_{r_1, r_2, \dots,r_m,r}}{m \Omega^{m-1}} \Gamma_{j_1i}^{r_1} \Gamma_{j_2i}^{r_1} \dots \Gamma_{j_mi}^{r_1} \right) \left( k_i^{r} R_i^r(\mathbf{s}) \right)    \right) \nonumber \\  
    +&  J^m \left(\sum_i \sum_j \sum_r  \left( M^{m}_{j_1^{r_1}, j_2^{r_2}, \dots j_{m-1}^{r_{m-1}},j^r} {\Gamma_{j_mi}^{r_m}\delta_{r_m,r}} +  M^{m-1}_{j_1^{r_1}, j_2^{r_2}, \dots j_{m-2}^{r_{m-2}},j^r} \frac{\Gamma_{j_{m-1}i}^{r_m} \Gamma_{j_mi}^{r_m}\delta_{r_{m-1},r_m,r}}{\Omega}  + \dots + \frac{\delta_{r_1, r_2, \dots,r_m,r}}{\Omega^{m-1}} \Gamma_{j_1i}^{r_1} \Gamma_{j_2i}^{r_1} \dots \Gamma_{j_{m-1}i}^{r_1} M^2_{j_m^{r_1}, j^r}  \right) \left( k_i^{r}   \frac{\partial R_i^r(\mathbf{s})}{\partial s_j^r} \right)  \right), 
\end{align}}

this result demonstrates that until first-order reactions occur, the temporal derivation of the $m$-th moment depends only on the $m$-th moment. Following a method similar to the diffusion part, Proposition 1 is proved.   

\section{Prof of Proposition 5 \label{B}} 
\textbf{Proof.:}\\    
First, we have $\alpha_{ij}^r \leq 1$, yielding the next propensity rates  
\begin{align}  
a_i^r(\mathbf{S}^r) &= k_i^{r} \prod_{j=1}^{N} \frac{S_j^r !}{\Omega^{\alpha_{ij}}(S_j^r - \alpha_{ij})!} = k_i^{r}  \prod_{j=1}^{N} \left( \frac{S_j^r}{\Omega} \right)^{\alpha_{ij}^r},
\end{align}  
from which we obtain the deterministic rates exactly as 
\begin{align}     
R_i^{r}(\mathbf{s}^r)= \frac{a_i^r(\braket{\mathbf{S}^r})}{k_i^{r}}= \prod_{j=1}^{N} (s_j^r)^{\alpha_{ij}^r}.  
\end{align} 
It is important to observe that each mean concentration $s_j^r$ appears up to exponent 1 because $\alpha_{ij}^r \leq 1$, and the second derivations according to the following conditions are 
\begin{align}  
\frac{\partial^m R_i^{r}(\mathbf{s}^r)}{ \partial s_{j_1}^{r} \partial s_{j_2}^{r}... \partial s_{j_m}^{r}} =  
\begin{cases}         
\neq  0 & \text{if } m=1 \text{ or }j_1 \neq j_2 \neq ... \neq j_m  \\         0 & \text{other cases}    
\end{cases}  , \label{b.3} 
\end{align} 
However, because all variables are independent, the covariance between the chemical species in the stationary state follows the following conditions uncorrelated:
\begin{itemize}
    \item  In the same voxel $r$: \\
      \begin{align}     
      M^{m}_{j_1^{r},j_2^{r}, ... , j_m^{r} }=   
      \begin{cases}         
       \neq 0 & \text{if } m \geq 2 \text{ and } j_1 = j_2= ...=j_m \\
       0 & \text{other cases}    \label{b.4}  
\end{cases} . 
\end{align}
    \item For two arbitrates voxels $r_1$ and $r_2$:\\
    {\small
    \begin{align}     
     M^{m+1}_{j_1^{r_1},j_2^{r_1}, ... , j_m^{r_1},l_2^{r_2} }=      
     \begin{cases}         
      \neq 0 & \text{if } m \geq 1 \text{ and } j_1 = j_2= ...=j_m=l_2 \\ 
        0 & \text{other cases}      
     \end{cases} . \label{b.5}
\end{align}} \\ 
\end{itemize}
now considering Equations (\ref{b.3})-(\ref{b.5}), we get 
\begin{align}
    M^m_{j_1^{r},j_2^{r}, ... , j_m^{r}}  \frac{\partial^m R_i^{r}(\mathbf{s}^r)}{ \partial s_{j_1}^{r} \partial s_{j_2}^{r}... \partial s_{j_m}^{r}}  &= 0 \text{ for any } m, \nonumber \\
    M^{m+1}_{j_1^{r_1},j_2^{r_1}, ... , j_m^{r_1},l_2^{r_2}}  \frac{\partial^m R_i^{r}(\mathbf{s}^r)}{ \partial s_{j_1}^{r_1} \partial s_{j_2}^{r_1}... \partial s_{j_m}^{r_1}}   &= 0 \text{ if } m\geq 2, \label{b.6}
\end{align}

Because the reaction network is a steady state, then the temporal derivatives in Equations (\ref{13}) and (\ref{14}) are zero, and considering Equations (\ref{b.6}), we get

\begin{align}     
\frac{\partial s_l^r }{\partial{t}}= 0 & =  \sum_{i} \Gamma_{li}^r  k_i^{r} R_i^{r}(\mathbf{s}^r)  +  \sum_q  \left(  d_{qr}^{l} {s_l^q} - d_{rq}^{l} {s_l^r} \right), \nonumber \\
\frac{\partial M^2_{{l}^{r_1}, {l}^{r_2}}}{\partial{t}}= 0  & = \sum_{i} \left( \delta_{r_1,r_2} \frac{\Gamma_{l i}^{r_1} \Gamma_{l i}^{r_1}}{\Omega} k_i^{r_1}R_i^{r_1}(\mathbf{s}^{r_1})  \right. \nonumber \\ 
& +  \left.  \left( M^2_{{l}^{r_1}, {l}^{r_2}} \Gamma_{l i}^{r_2}  k_i^{r_2} \frac{\partial R_i^{r_2}(\mathbf{s}^{r_2})}   {\partial {s_{l}^{r_2}}}  + M^2_{{l}^{r_1}, {l}^{r_2}} \Gamma_{l i}^{r_1}k_i^{r_1} \frac{\partial R_i^{r_1}(\mathbf{s}^{r_1})}{\partial {s_{l}^{r_1}}} \right)  \right) \nonumber \\
&+  \frac{1}{\Omega} \left( \delta_{r_1,r_2} \sum_q \left( d_{q r_1}^{l} s_{l}^{q} + d_{r_1 q}^{l} s_{l}^{r_1} \right) -(d_{r_1 r_2}^{l} s_{l}^{r_1} + d_{r_2 r_1}^{l} s_{l}^{r_2}) \right) \nonumber \\
&+  \sum_q \left( d_{q{r_1}}^{l} M^2_{l^{q},l^{r_2}}  - d_{{r_1}q}^{l} M^2_{l^{r_1},l^{r_2}} \right) +  \sum_q \left( d_{sr_2}^{l} M^2_{l^{r_1},l^{q}}  -d_{r_2s}^{l} M^2_{l^{r_1},l^{r_2}}  \right),  
\end{align}

where $R_i^{r}(\mathbf{s}^{r})= \prod_j (s_j^r)^{\alpha_{ij}^r}$. $\qed $

This result confirms Proposition 5.

\section{Hill Function \label{C}}
In this section, we derive the Hill function, and we base this derivation on \cite{Manuel, Decimal}. For this, we suppose that we have the following reactions 
\begin{eqnarray}
    R+nL \stackbin[k_{-}]{k_{+}}{\rightleftarrows} RL_n, \nonumber \\
    0 \stackbin[k_1]{k_2}{\rightleftarrows} L,
\end{eqnarray}
The first reaction is the binding of $n$ ligands $L$ to receptor $R$ in a reversible process to form complex $RL_n$. We purposely wrote the last reaction because $L$ may be subject to other reactions, and this reaction is a birth-death process that synthesizes and degrades ligands. The stoichiometric coefficients and the stoichiometric matrix are, 

\begin{align}
 \alpha_{ij}&= \begin{pmatrix}
		n& 1 & 0 \\
            0& 0 & 1  \\
            0& 0 & 0  \\
            1& 0 & 0
	\end{pmatrix} ,    &
 \beta_{ij}&=  \begin{pmatrix}
		0& 0 & 1 \\
            n& 1 & 0  \\
            1& 0 & 0  \\
            0& 0 & 0 
	\end{pmatrix}, \nonumber \\
 \Gamma_{ij}&= \begin{pmatrix}
		-n & n & 1 & -1 \\
		-1 & 1 & 0 & 0 \\
		1  & -1 & 0 & 0 
	\end{pmatrix}. \label{c.1}
\end{align}
Let $L, R, S$ be the number of molecules of $L, R, RL_n$ respectively. Thus, the propensity rates for getting the Hill functions are
\begin{align}
    a_1= k_{+} R \frac{L!}{(L-n)!} \frac{1}{\Omega^{n+1}}, & a_2=& k_{-} S\frac{1}{\Omega},     \label{c.2}
\end{align}
there is a conservative quantity $R+S=R_0$, the number of initial receptors, then the previous equations are reduced to 
\begin{align}
    a_1&= k_{+} (R_0-S) \frac{L!}{(L-n)!} \frac{1}{\Omega^{n+1}}, & a_2=& k_{-} S\frac{1}{\Omega},     \label{c.3}
\end{align}
and 
\begin{align}
     \Gamma'_{ij}&= \begin{pmatrix}
		-n & n & 1 & -1 \\
		1  & -1 & 0 & 0 
	\end{pmatrix},  
\end{align}
if we suppose that the first reactions in (\ref{c.1}) are in the stationary state, $S$ and $L$ are independent, then we can use Proposition 3, where the central moments between $S$ and $L$ are zero, then we get the next equation

\begin{align}
    \frac{\partial s}{\partial t}=0 = -k_{-} s + k_{+}(r_0 - s) \frac{1}{\Omega^{n}} \left \langle   \frac{L!}{(L-n)!}  \right \rangle,
\end{align}
where $K^n= \frac{k_{-}}{k_{+}}$,  $r_0=R_0/\Omega$, $s=\braket{S}/\Omega$ and $r=\braket{R}/\Omega$ are the mean concentrations of  $S$ and $R$, from this we get 
\begin{align}
    s= r_0 \frac{\frac{1}{\Omega^{n}}\left \langle   \frac{L!}{(L-n)!}  \right \rangle}{K^n+ \frac{1}{ \Omega^{n}}\left \langle   \frac{L!}{(L-n)!}  \right \rangle}.
\end{align}
If we define the Hill function as follows and substitute $r+s=r_0$ and the value of $s$, we have
\begin{align}
    H=\frac{s}{r+s}= \frac{s}{r_0} = \frac{\frac{1}{\Omega^{n}}\left \langle   \frac{L!}{(L-n)!}  \right \rangle}{K^n+ \frac{1}{ \Omega^{n}}\left \langle   \frac{L!}{(L-n)!}  \right \rangle}, \label{c.5}
\end{align}
now we show some forms of this for some specific cases, we used $l= \braket{L}/\Omega$ and $M^m_{l, l,...,l}= \braket{(L-\braket{L})^m}/\Omega^m$, 
\begin{itemize}
    \item If $n=1$: 
    \begin{align}
        H= \frac{l}{K+ l}.
    \end{align}
    \item If $n=2$:
    \begin{align}
        H= \frac{l^2 + M^2_{l,l} - \frac{l}{\Omega}  }{{K^2} + l^2 + M^2_{l,l} - \frac{l}{\Omega} }.
    \end{align}
    \item If $n=3$:
    \begin{align}
        H= \frac{ (l^3 + 3l M^2_{l,l} + M^3_{l,l,l}) - \frac{3}{\Omega}(l^2 + M^2_{l,l} ) + \frac{2 l }{\Omega^2}}{K^3+ (l^3 + 3l M^2_{l,l} + M^3_{l,l,l}) - \frac{3}{\Omega}(l^2 + M^2_{l,l} ) + \frac{2 l }{\Omega^2} }.
    \end{align}
\end{itemize}
These expressions are exact because we did not make any approximations. All of these equations are for activators, but we can do some similar ones for repressors or only use the relation $D=1-H$.

If, in the stationary state, the distribution of $L$ follows a Poisson distribution given by  
\begin{align}  
P_{ss}(L) = e^{-\braket{L}_{ss}} \frac{\braket{L}_{ss}^L}{L!}, 
\end{align}  
where $\braket{L}_{ss}$ is the mean in the stationary state, and evaluating the Hill function (\ref{c.5}) in the stationary state yields  
\begin{align}    
H = \frac{\frac{\braket{L}_{ss}^n}{\Omega^n}}{K^n + \frac{\braket{L}_{ss}^n}{\Omega^n}} = \frac{l_{ss}^n}{K^n + l_{ss}^n}. 
\end{align}   
This result demonstrates that, in the stationary state, the Hill function with stochastic corrections is identical to the deterministic Hill function.

\section{Non Converge for Oscillations \label{F}}

\begin{figure*} [h!t]
  \begin{subfigure}{\linewidth}
\includegraphics[width=.24\textwidth]{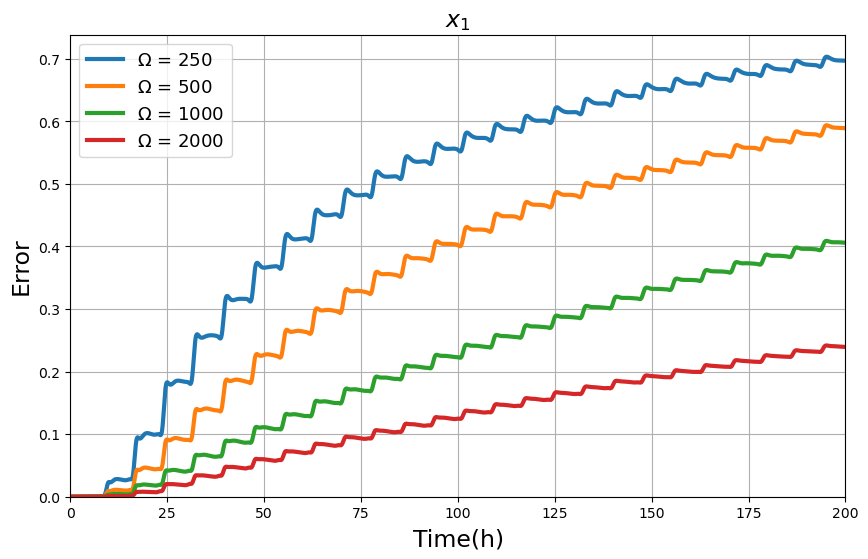}\hfill
\includegraphics[width=.24\textwidth]{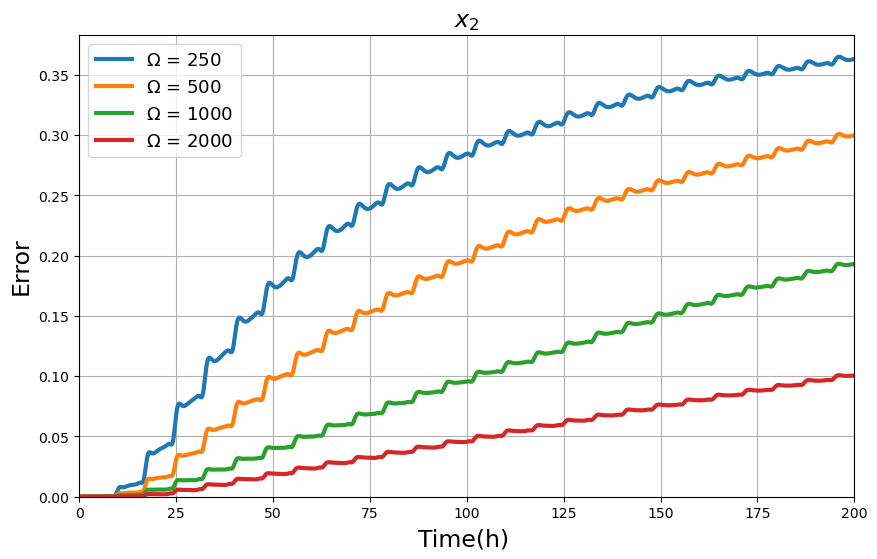}\hfill
\includegraphics[width=.24\textwidth]{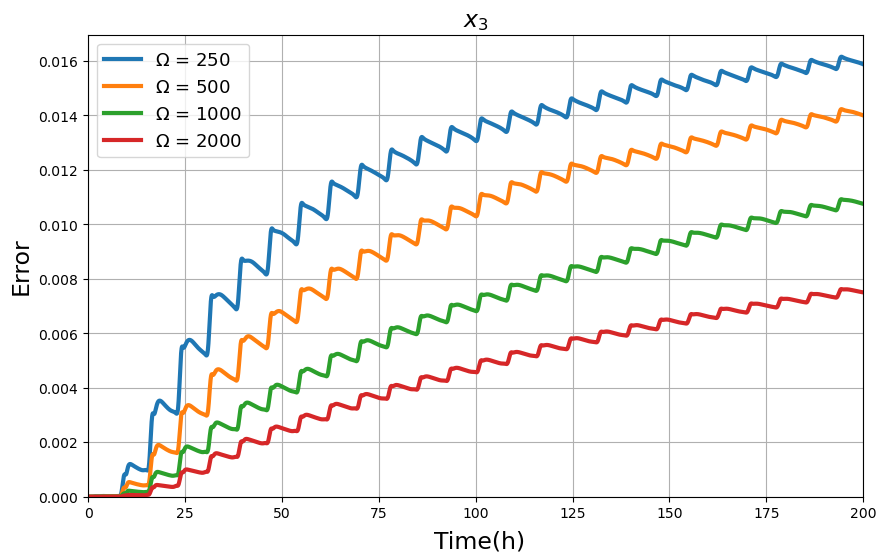}\hfill
\includegraphics[width=.24\textwidth]{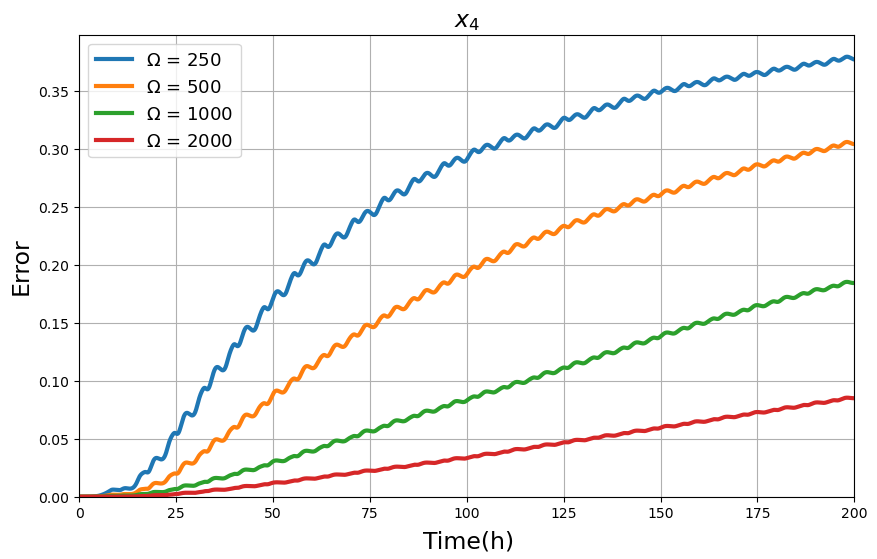}
\caption{\textbf{Mean concentration} }
  \end{subfigure}\par\medskip
    \begin{subfigure}{\linewidth}
\includegraphics[width=.24\textwidth]{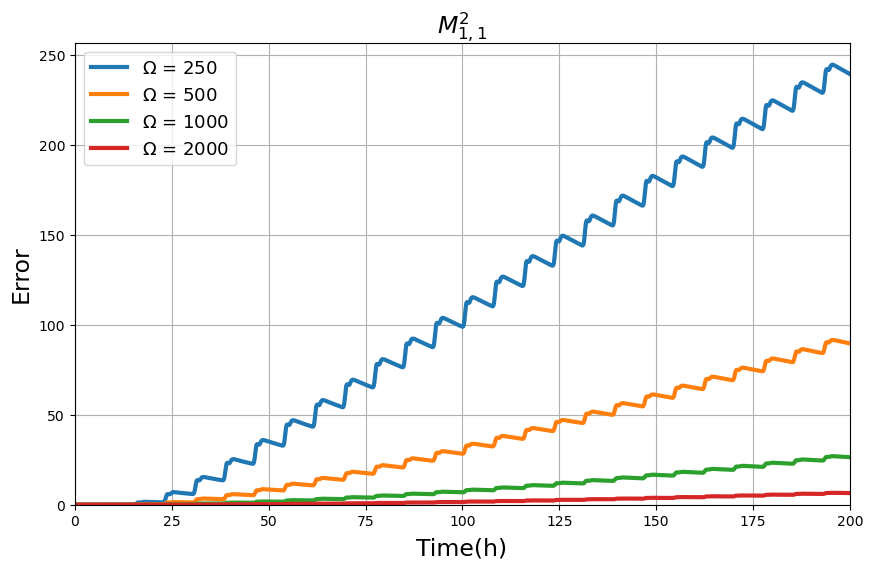}\hfill
\includegraphics[width=.24\textwidth]{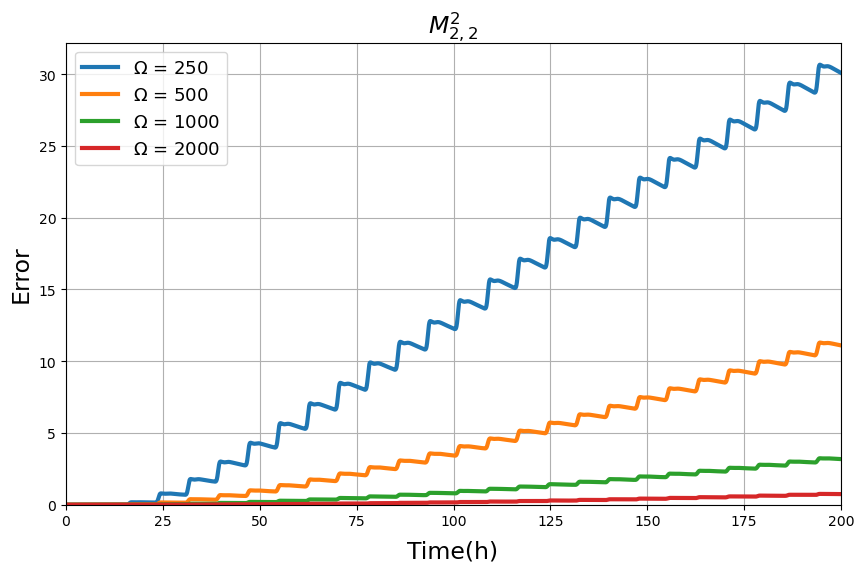}\hfill
\includegraphics[width=.24\textwidth]{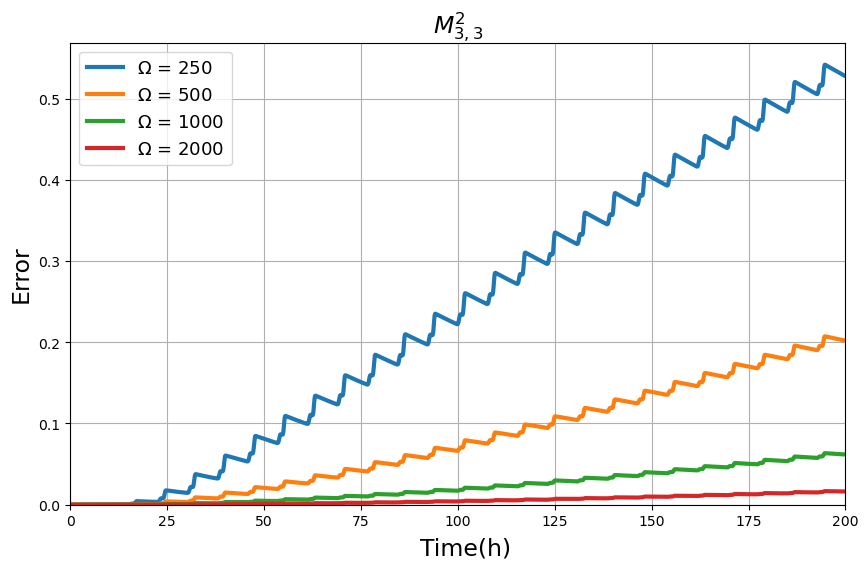}\hfill
\includegraphics[width=.24\textwidth]{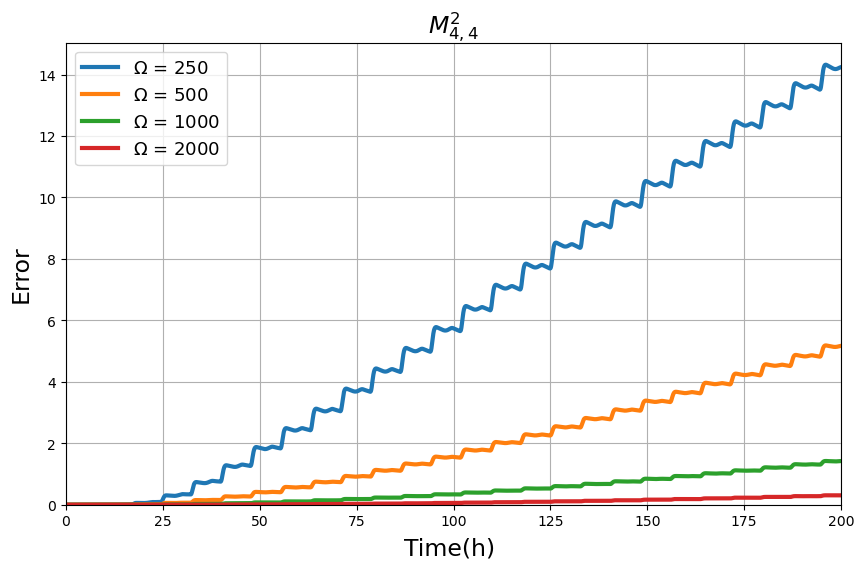}
\caption{\textbf{Second Central Moment} }
  \end{subfigure}
\caption{\textbf{Error in Antithetic with Oscillations.} In these figures, we graphed the error at each instant in time according to Equation (\ref{15}). In panel a), we calculate the error of the mean of the variables, and in panel b), the variance of the same variables.  We can see that, when the size of the system $\Omega$ increases, the error decreases with respect to the others. However, in this case, the error increases for all means and variances. The parameters and initial conditions are listed in Table \ref{tabla20}. }
\label{fig.f1}
\end{figure*}

We repeated the same analysis for the antithetic system and quantified the error at each time point by using Equation (\ref{20}). The parameters were selected such that the system exhibited oscillatory behavior (see Table \ref{tabla20}). The analysis revealed that the error increases, resulting in poor approximation, as shown in Figure \ref{fig.f1}. This is because the oscillations are lost in the stochastic simulations owing to variations in the amplitude and period across individual realizations. Consequently, the moment-closure approach is unsuitable for systems with high-order reactions that oscillate. Further investigation is required to address this limitation.

\section{Parameters}

Tables \ref{tabla1}–\ref{tabla4} list the parameters and initial conditions for each model presented in this study. 

\begin{table*}[htbp]
\centering
\begin{tabular}{|p{2.6cm}|c|c|c|}
\hline
\textbf{Parameters} & \textbf{Description} & \textbf{Value} \\
\hline
$k_1$ & $x_2$ Synthesis rate. & 1 $h^{-1}$ \\
$k_2$ & $x_2$ Degradation rate. & 1 $h^{-1}$ \\
$k_3$ & Union rate of $x_1$ and $x_2$. & 1 $(mol\times h)^{-1}$. \\
$k_4$ & Separation rate of $x_3$. & 1 $h^{-1}$\\
$k_5$ & Production rate of $x_4$.  & 1 $h^{-1}$ \\
$k_6$ &  Union rate of $x_2$ and $x_4$. & 1 $h^{-1}$ \\
$x_1(0)$& Initial mean concentration of $x_1$. & 1 $mol$ \\
$x_2(0)$& Initial mean concentration of $x_2$. & 1 $mol$ \\
$x_3(0)$& Initial mean concentration of $x_3$. & 0 $mol$ \\
$x_4(0)$& Initial mean concentration of $x_4$. & 0 $mol$ \\
$M^2_{i,j}(0)$& Initial second central moment of concentration of species. & 0 $mol^2$ \\
\hline
\end{tabular}
\caption{\textbf{Parameters and initial conditions for the enzyme process.} ($i,j \in$ {$1,2,3,4$}) In this table, we show the parameters and initial conditions used for the system.}
\label{tabla1}
\end{table*}

\begin{table*}[htbp]
\centering
\begin{tabular}{|p{2.6cm}|c|c|c|}
\hline
\textbf{Parameters} & \textbf{Description} & \textbf{Value} \\
\hline
$k_1$ & Synthesis rate of $x_2$ by $x_1$. & 1 $(mol \times h)^{-1}$ \\
$\theta_2$ & Synthesis rate of $x_4$ by $x_2$. & 1 $(mol \times h)^{-1}$  \\
$\theta_1$ & Synthesis rate of $x_1$ by $x_3$. & 1 $(mol \times h)^{-1}$  \\
$\eta$ & Degradation rate of the complex formed by $x_3$ by $x_4$. & 10 $(mol \times h)^{-1}$  \\
$\gamma_p$ & Degradation rate of $x_1$ and $x_2$. & 1 $h^{-1}$ \\
$\mu$ & Synthesis rate of $x_3$. & 10 $h^{-1}$\\
$x_1(0)$& Initial mean concentration of $x_1$. & 1 $mol$ \\
$x_2(0)$& Initial mean concentration of $x_2$. & 1 $mol$ \\
$x_3(0)$& Initial mean concentration of $x_3$. & 0.1 $mol$ \\
$x_4(0)$& Initial mean concentration of $x_4$. & 1 $mol$ \\
$M^2_{i,j}(0)$& Initial second central moment of concentration of species. & 0 $mol^2$ \\
\hline
\end{tabular}
\caption{\textbf{Parameters and initial conditions for the antithetic with a stationary point.} ($i,j \in$ {$1,2,3,4$}) In this table, we show the parameters and initial conditions used for the system when this has a stationary point; these parameters were obtained from  \cite{Briat}. }
\label{tabla2}
\end{table*}

\begin{table*}[htbp]
\centering
\begin{tabular}{|p{2.6cm}|c|c|c|}
\hline
\textbf{Parameters} & \textbf{Description} & \textbf{Value} \\
\hline
$k_1$ & Synthesis rate of $x_2$ by $x_1$. & 1 $(mol \times h)^{-1}$ \\
$\theta_2$ & Synthesis rate of $x_4$ by $x_2$. & 1 $(mol \times h)^{-1}$  \\
$\theta_1$ & Synthesis rate of $x_1$ by $x_3$. & 1 $(mol \times h)^{-1}$  \\
$\eta$ & Degradation rate of the complex formed by $x_3$ by $x_4$. & 40 $(mol \times h)^{-1}$  \\
$\gamma_p$ & degradation rate of $x_1$ and $x_2$. & 1 $h^{-1}$ \\
$\mu$ & Synthesis rate of $x_3$. & 10 $h^{-1}$\\
$x_1(0)$& Initial mean concentration of $x_1$. & 1 $mol$ \\
$x_2(0)$& Initial mean concentration of $x_2$. & 1 $mol$ \\
$x_3(0)$& Initial mean concentration of $x_3$. & 0.1 $mol$ \\
$x_4(0)$& Initial mean concentration of $x_4$. & 1 $mol$ \\
$M^2_{i,j}(0)$& Initial second central moment of concentration of species. & 0 $mol^2$ \\
\hline
\end{tabular}
\caption{\textbf{Parameters and initial conditions for the antithetic with a stable limit cycle.} ($i,j \in$ {$1,2,3,4$}) In this table, we show the parameters and initial conditions used for the system when this has a stable limit cycle; these parameters were obtained from  \cite{Briat}. }
\label{tabla20}
\end{table*}

\begin{table*}[htbp]
\centering
\begin{tabular}{|p{2.6cm}|c|c|c|}
\hline
\textbf{Parameters} & \textbf{Description} & \textbf{Value} \\
\hline
$k_1$ & $x_1$ Synthesis rate. & 10 $h^{-1}$ \\
$k_2$ & $x_1$ Degradation rate. & 0.5 $h^{-1}$ \\
$k_3$ & $x_2$ Synthesis rate. & 10 $h^{-1}$ \\
$k_4$ & $x_2$ Degradation rate. & 0.5 $h^{-1}$\\
$x_1(0)$& Initial mean concentration of $x_1$. & 0 $mol$ \\
$x_2(0)$& Initial mean concentration of $x_2$. & 8 $mol$ \\
$M^2_{i,i}(0)$& Initial second central moment of concentration of species. & 0 $mol^2$ \\
$M^3_{i,i,i}(0)$& Initial third central moment of concentration of species. & 0 $mol^3$ \\
$\Omega$& Size of the system. & 100 $(mol)^{-1}$ \\
\hline
\end{tabular}
\caption{\textbf{Parameters and initial conditions for a genetic network with negative feedback without diffusion.} ($i\in$ {$1,2$}) In this table, we show the parameters and initial conditions used for the system.}
\label{tabla3}
\end{table*}

\begin{table*}[htbp]
\centering
\begin{tabular}{|p{2.6cm}|c|c|c|}
\hline
\textbf{Parameters} & \textbf{Description} & \textbf{Value} \\
\hline
$k_1$ & $x_1$ Synthesis rate. & 10 $h^{-1}$ \\
$k_2$ & $x_1$ Degradation rate. & 0.5 $h^{-1}$ \\
$k_3$ & $x_2$ Synthesis rate. & 10 $h^{-1}$ \\
$k_4$ & $x_2$ Degradation rate. & 0.5 $h^{-1}$\\
$x_1^r(0)$& Initial mean concentration of $x_1$. & (random) $mol$ \\
$x_2^r(0)$& Initial mean concentration of $x_2$. & (random) $mol$ \\
$M^2_{i^{r_1},i^{r_2}}(0)$& Initial second central moment of concentration of species. & 0 $mol^2$ \\
$\Omega$& Size of the system. & 100 $(mol)^{-1}$ \\
$D_1$& Diffusion rate of $x_1$. & 1 $(mm)^2$ \\
$D_2$& Diffusion rate of $x_2$. & 0.5 $(mm)^2$ \\
$\lambda$& Size of the voxel. & 1 $(mm)$ \\
\hline
\end{tabular}
\caption{\textbf{Parameters and initial conditions for a genetic network with negative feedback with diffusion.} ($i\in$ {$1,2$}) In this table, we show the parameters and initial conditions used for the system.}
\label{tabla4}
\end{table*}

\end{document}